%% file: LogicalMajoranaFermions_arXiv.tex
\documentclass[pra,twocolumn,superscriptaddress,nofootinbib]{revtex4}

\pdfoutput=1

\newcommand{\FigDirectory}{figs}

\input{UsePackages}

\input{landahl_newcommands}

\begin{document}

    %%%%%%%%%%%%%%%%%%%%%%%%%%%%%%%%%%%%%%%%%%%%%%%%%%%%%%%%%%%%%%%%%%%%%%%%%%%%
    % Title & Authors
    % For PRA, this comes after \begin{document}.

    \title{Logical fermions for fault-tolerant quantum simulation}

    \author{Andrew J. \surname{Landahl}}
    \email[]{alandahl@sandia.gov}
    \affiliation{Center for Computing Research,
                 Sandia National Laboratories,
                 Albuquerque, NM, 87185, USA}
    \affiliation{Center for Quantum Information and Control,
                 University of New Mexico,
                 Albuquerque, NM, 87131, USA}
    \affiliation{Department of Physics and Astronomy,
                 University of New Mexico,
                 Albuquerque, NM, 87131, USA}
    \author{Benjamin C.~A. \surname{Morrison}}
    \email[]{benmorrison@unm.edu}
    \affiliation{Center for Computing Research,
                 Sandia National Laboratories,
                 Albuquerque, NM, 87185, USA}
    \affiliation{Center for Quantum Information and Control,
                 University of New Mexico,
                 Albuquerque, NM, 87131, USA}
    \affiliation{Department of Physics and Astronomy,
                 University of New Mexico,
                 Albuquerque, NM, 87131, USA}

    %%%%%%%%%%%%%%%%%%%%%%%%%%%%%%%%%%%%%%%%%%%%%%%%%%%%%%%%%%%%%%%%%%%%%%%%%%%%
    % Abstract

    \begin{abstract}
	\input{Abstract}
    \end{abstract}

    %\date[\bf DRAFT:\rm\ ]{\DTMnow}

    % For a PRA, \maketitle comes after the abstract.
    \maketitle

    %%%%%%%%%%%%%%%%%%%%%%%%%%%%%%%%%%%%%%%%%%%%%%%%%%%%%%%%%%%%%%%%%%%%%%%%%%%%
    % Body

    \input{Body}

	\input{Conclusion}

    %%%%%%%%%%%%%%%%%%%%%%%%%%%%%%%%%%%%%%%%%%%%%%%%%%%%%%%%%%%%%%%%%%%%%%%%%%%%
    % Acknowledgment

    \begin{acknowledgments}
    \vspace{5ex}
    \input{Acknowledgments}
    \end{acknowledgments}

    %%%%%%%%%%%%%%%%%%%%%%%%%%%%%%%%%%%%%%%%%%%%%%%%%%%%%%%%%%%%%%%%%%%%%%%%%%%%
    % References

    % If hyperref is included, then \phantomsection is already defined.
    % If not, we need to define it.
    \providecommand*{\phantomsection}{}
    \phantomsection
    %\addcontentsline{toc}{section}{References}
    %\bibliographystyle{plain}
    %\bibliographystyle{landahlamsalpha}
    \bibliographystyle{landahl}
    
    \bibliography{landahl,morrison}

    %%%%%%%%%%%%%%%%%%%%%%%%%%%%%%%%%%%%%%%%%%%%%%%%%%%%%%%%%%%%%%%%%%%%%%%%%%%%
    % Appendix

    \appendix
    \input{Appendix}

\end{document}

%% file: UsePackages.tex
\usepackage{dsfont}                  % doublestroke \CC, etc.
\usepackage[usenames]{color}         % font colors.
\usepackage[OT2,T1]{fontenc}         % To allow cyrillic later

\usepackage[FIGBOTCAP,normal,bf,tight]{subfigure}
\usepackage{float}                   % to force figure locations [H]
\usepackage{graphicx}                % ps, pdf figs.
\usepackage{amsmath}           % Better math control
\usepackage{amsfonts}          % Extended math fonts
\usepackage{amssymb}           % Names amsfont symbols
\usepackage[colorlinks=true,
            urlcolor=blue,
            citecolor=blue,
            linkcolor=black,
            pdfstartview=FitH,
            pdfpagemode=None]{hyperref}

\usepackage{booktabs}         % for tables
\usepackage{tabulary}         % like tabularx, but better
\usepackage{multirow}
\usepackage[en-US]{datetime2}
\usepackage{comment}
\usepackage{tikz}
\usetikzlibrary{quantikz}

%% file: landahl_newcommands.tex
\newcommand{\ssc}[2][]{^{\vphantom\dagger{#1}}_{#2}}

\newcommand{\ZZ}{{\mathds{Z}}}

\newcommand{\RR}{{\mathds{R}}}

\renewcommand{\>}{\rangle}

\newcommand{\lsim}{\mathrel{\raise.3ex\hbox{$<$\kern-.75em\lower1ex\hbox{$\sim$}}}}
\newcommand{\gsim}{\mathrel{\raise.3ex\hbox{$>$\kern-.75em\lower1ex\hbox{$\sim$}}}}

\def\QECCnk[[#1,#2]]{[\![#1, #2]\!]}

\def\QECCnkq[[#1,#2,#3]]{[\![#1, #2]\!]_{#3}^{\vphantom{T}}}

\def\QECCnkd[[#1,#2,#3]]{[\![#1, #2, #3]\!]}

\def\QECCnkdq[[#1,#2,#3,#4]]{[\![#1, #2, #3]\!]_{#4}^{\vphantom{T}}}

\def\QECCnkgd[[#1,#2,#3,#4]]{[\![#1, #2, #3, #4]\!]}

\def\QECCnkgdq[[#1,#2,#3,#4,#5]]{[\![#1, #2, #3, #4]\!]_{#5}^{\vphantom{T}}}

\def\QECCnkdc[[#1,#2,#3,#4]]{[\![#1, #2, #3; #4]\!]}

\def\QECCnkdcq[[#1,#2,#3,#4,#5]]{[\![#1, #2, #3; #4]\!]_{#5}^{\vphantom{T}}}

\def\QECCnkgdcq[[#1,#2,#3,#4,#5,#6]]{%
  [\![#1, #2, #3, #4; #5]\!]_{#6}^{\vphantom{T}}}

\newcommand{\bigO}{{\cal O}}

\def\openone{\leavevmode\hbox{\small1\normalsize\kern-.33em1}}

\newcommand{\etal}{\textit{et~al.}}
\newcommand{\viz}{\textit{viz.}}

\newcommand{\eg}{\textit{e.g.}}

\long\def\symbolfootnote[#1]#2{\begingroup%
\def\thefootnote{\fnsymbol{footnote}}\footnote[#1]{#2}\endgroup}

%% file: Abstract.tex
%%%%%%%%%%%%%%%%%%%%%%%%%%%%%%%%%%%%%%%%%%%%%%%%%%%%%%%%%%%%%%%%%%%%%%%%%%%%%%%%
% File: Abstract.tex
%
% Authors: Andrew J. Landahl <alandahl@sandia.gov>
%          Benjamin C. A. Morrison <benmorr@sandia.gov>
%
%%%%%%%%%%%%%%%%%%%%%%%%%%%%%%%%%%%%%%%%%%%%%%%%%%%%%%%%%%%%%%%%%%%%%%%%%%%%%%%%

%%%%%%%%%%%%%%%%%%%%%%%%%%%%%%%%%%%%%%%%%%%%%%%%%%%%%%%%%%%%%%%%%%%%%%%%%%%%
% Abstract

We show how to absorb fermionic quantum simulation's expensive fermion-to-qubit
mapping overhead into the overhead already incurred by surface-code-based
fault-tolerant quantum computing.  The key idea is to process information in
surface-code twist defects, which behave like logical Majorana fermions.  Our
approach encodes Dirac fermions, a key data type for simulation applications,
directly into logical Majorana fermions rather than atop a logical qubit layer
in the architecture. Using quantum simulation of the $N$-fermion 2D
Fermi-Hubbard model as an exemplar, we demonstrate two immediate algorithmic
improvements. First, by preserving the model's locality at the logical level, we
reduce the asymptotic Trotter-Suzuki quantum circuit depth from
$\bigO(\sqrt{N})$ in a typical Jordan-Wigner encoding to $\bigO(1)$ in our
encoding. Second, by exploiting optimizations manifest for logical fermions but
less obvious for logical qubits, we reduce the $T$-count of the block-encoding
\textsc{select} oracle by 20\% over standard implementations, even when realized by
logical qubits and not logical fermions.

%% file: Body.tex
%%%%%%%%%%%%%%%%%%%%%%%%%%%%%%%%%%%%%%%%%%%%%%%%%%%%%%%%%%%%%%%%%%%%%%%%%%%%%%%%
% File: Body.tex
%
% Authors: Andrew J. Landahl <alandahl@sandia.gov>
%          Benjamin C. A. Morrison <benmorr@sandia.gov>
%
%%%%%%%%%%%%%%%%%%%%%%%%%%%%%%%%%%%%%%%%%%%%%%%%%%%%%%%%%%%%%%%%%%%%%%%%%%%%%%%%

%%%%%%%%%%%%%%%%%%%%%%%%%%%%%%%%%%%%%%%%%%%%%%%%%%%%%%%%%%%%%%%%%%%%%%%%%%%%%%%
% Introduction

\input{Introduction}

%%%%%%%%%%%%%%%%%%%%%%%%%%%%%%%%%%%%%%%%%%%%%%%%%%%%%%%%%%%%%%%%%%%%%%%%%%%%%%%
% Background

\input{Background}

%%%%%%%%%%%%%%%%%%%%%%%%%%%%%%%%%%%%%%%%%%%%%%%%%%%%%%%%%%%%%%%%%%%%%%%%%%%%%%%
% LogicalOps

\input{Results}

%%%%%%%%%%%%%%%%%%%%%%%%%%%%%%%%%%%%%%%%%%%%%%%%%%%%%%%%%%%%%%%%%%%%%%%%%%%%%%%
% NISQ Testbed

%\input{NISQTestbed}

%% file: Introduction.tex
%%%%%%%%%%%%%%%%%%%%%%%%%%%%%%%%%%%%%%%%%%%%%%%%%%%%%%%%%%%%%%%%%%%%%%%%%%%%%%%%
% File: Introduction.tex
%
% Authors: Andrew J. Landahl <alandahl@sandia.gov>
%          Benjamin C. A. Morrison <benmorr@sandia.gov>
%
%%%%%%%%%%%%%%%%%%%%%%%%%%%%%%%%%%%%%%%%%%%%%%%%%%%%%%%%%%%%%%%%%%%%%%%%%%%%%%%%

%%%%%%%%%%%%%%%%%%%%%%%%%%%%%%%%%%%%%%%%%%%%%%%%%%%%%%%%%%%%%%%%%%%%%%%%%%%%%%%
% Section
%
\section{Introduction}
\label{sec:introduction}

In 1982, Richard Feynman famously argued that if we plan to simulate quantum
mechanics with computers, then we should use \emph{quantum} computers to do so
\cite{Feynman:1982a}.  He was right---quantum computers are expected to excel at
simulating quantum systems, taking us far beyond the reach of conventional
computers \cite{Lloyd:1996a, Georgescu:2014a}.  In fields like quantum
chemistry, materials science, nuclear physics, and high-energy physics, these
systems are frequently comprised of \emph{fermions} \cite{ASCR:2015a, HEP:2015a,
BES:2017a, NP:2018a, FES:2018a, Bauer:2019a, NSAC-QIS:2019a}.  If the quantum
computer simulating them uses \emph{qubits} to store quantum information, then
there is an unavoidable and substantial overhead required to map the fermions to
the qubits \cite{Jiang:2020a}.  Examples of such mappings include the
Jordan-Wigner transformation \cite{Jordan:1928a}, the Verstraete-Cirac
transformation \cite{Verstraete:2005a}, the Bravyi-Kitaev transformation
\cite{Bravyi:2000a, Seeley:2012a} (and its ``superfast'' implementation via
Fenwick trees \cite{Havlicek:2017a}), and the ternary tree transformation
\cite{Jiang:2020a}; see also Refs.~\cite{Jiang:2018a, Steudtner:2019a,
Setia:2019a, Derby:2020a, Kirby:2021a}.  Moreover, these qubits
must have very low error rates for the results to be trustworthy.  This means
that the relevant qubits will likely need to be \emph{logical} (\viz, encoded)
qubits that are realized by quantum error correcting codes, and these codes must
be processed using fault-tolerant quantum computing protocols, which adds even
more overhead.

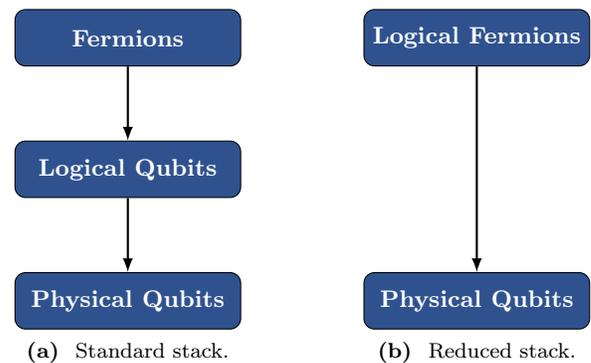
\begin{figure}[H]
\definecolor{sblue}{rgb}{.184,.322,.561}
\tikzstyle{box} = [rectangle, rounded corners, minimum width=3.0cm, minimum height=0.75cm, align=center, draw=black, fill=sblue, text=white]
\tikzstyle{arrow} = [thick,->,>=latex]
\tikzstyle{arrowlabel} = [align=center,fill=white]
\center{
  \subfigure[\
Standard stack.]{
\begin{tikzpicture}[node distance=1.75cm and 4cm, on grid]
\node (fermions) [box] {\textbf{Fermions}};
\node (logical_qubits) [box, below=of fermions] {\textbf{Logical Qubits}};
\node (physical_left) [box, below=of logical_qubits] {\textbf{Physical Qubits}};

\draw [arrow] (fermions) -- (logical_qubits);
\draw [arrow] (logical_qubits) --  (physical_left);
\end{tikzpicture}
}
\qquad\qquad
  \subfigure[\label{fig:lf-pqb}\
Reduced stack.]{
\begin{tikzpicture}[node distance=1.75cm and 4cm, on grid]
\node (logical_fermions) [box] {\textbf{Logical Fermions}};
\node (middle_right) [shape=rectangle, below=of logical_fermions] {};
\node (physical_right) [box, below=of middle_right] {\textbf{Physical Qubits}};

\draw [arrow] (logical_fermions) -- (physical_right);
\end{tikzpicture}
}
}
\caption{\small{\label{fig:fermion-stack}Two choices for the
fermion-to-qubit quantum software stack.}}
\end{figure}

Feynmanian thinking suggests that we should use \emph{fermionic} quantum
computers to simulate fermionic systems.  The fermionic quantum circuit model
developed by Bravyi and Kitaev fits the bill \cite{Bravyi:2000a}.  In this
model, elementary gates act as spacetime braids on the worldlines of Majorana
fermions; any ordinary (Dirac) fermion can always be mathematically split into
a pair of such Majorana fermions \cite{Majorana:1937a}.  Realizing this model
with physical Majorana zero modes has been a decades-long quest that is still
not complete, although progress continues to be made \cite{Aghaee:2022a}.  As an
alternative, one can instead use ``synthetic'' Majorana fermions constructed
from ordinary qubits arranged and processed with quantum surface codes
\cite{Bravyi:1998a, Dennis:2002a}.  In these codes, ``twist'' defects in the
lattice defining the codes act as \emph{logical} Majorana fermions that are
protected from noise \cite{Bombin:2010b, Brown:2017a, Kesselring:2018a}.  By
processing these logical Majorana fermions fault-tolerantly, there is an
opportunity to eliminate the expensive fermion-to-logical-qubit mapping in the
standard quantum software stack, as depicted in Fig.~\ref{fig:fermion-stack}.
In other words, it becomes possible to treat logical (Majorana) fermions as
elementary data types that can be processed directly in fermionic quantum
simulation algorithms.  Hybrid algorithms that process logical qubits alongside
logical fermions are also possible.

Embracing the logical (Majorana) fermion as an elementary data type for quantum
computers is akin to embracing the floating point number, or \emph{float}, as an
elementary data type for classical computers.  For machines that process qubits
or bits at the lowest level, fermions or floats might seem exotic.  However,
their existence belies the fact that many of the programs that we want to run on
these computers naturally process these data types.  Classical computers are
even typically benchmarked by how many floating point operations per second
(flops) they can achieve.  Perhaps someday quantum computers will be
benchmarked by how many fermionic operations per second they can achieve.

In this paper, we develop the reduced-stack architecture depicted in
Fig.~\ref{fig:lf-pqb} and show how it can be exploited to deliver computational
speedups for fermionic quantum simulation algorithms.

The remainder of this paper is organized as follows.

In Sec.~\ref{sec:background}, we review the pertinent background on
fermionic quantum circuits, logical fermion codes, and fermionic simulation
algorithms.
In Sec.~\ref{sec:logical-Majorana-operations}, we
describe how to realize a logical fermionic architecture stack, compare it to
the conventional error-corrected architecture for simulation, and examine
how it can provide performance improvements in two exemplar systems.
In Sec.~\ref{sec:conclusion}, we conclude.

%% file: Background.tex
%%%%%%%%%%%%%%%%%%%%%%%%%%%%%%%%%%%%%%%%%%%%%%%%%%%%%%%%%%%%%%%%%%%%%%%%%%%%%%%%
% File: Background.tex
%
% Authors: Andrew J. Landahl <alandahl@sandia.gov>
%          Benjamin C. A. Morrison <benmorr@sandia.gov>
%
%%%%%%%%%%%%%%%%%%%%%%%%%%%%%%%%%%%%%%%%%%%%%%%%%%%%%%%%%%%%%%%%%%%%%%%%%%%%%%%%

%%%%%%%%%%%%%%%%%%%%%%%%%%%%%%%%%%%%%%%%%%%%%%%%%%%%%%%%%%%%%%%%%%%%%%%%%%%%%%%
% Section
%
\section{Background}
\label{sec:background}

%%%%%%%%%%%%%%%%%%%%%%%%%%%%%%%%%%%%%%%%%%%%%%%%%%%%%%%%%%%%%%%%%%%%%
% Subsection
%
\subsection{Majorana fermions and fermionic Hamiltonians}
\label{sec:Majorana fermions}

Consider a collection of fermions obeying the following {canonical
anticommutation relations} on their elementary creation ($f^\dagger$) and
annihilation ($f$) operators for modes labeled by a non-negative integer $p$:
\begin{align}
\label{eq:CCR-fermions}
\{f\ssc{p}, f_q^\dagger\} &= \delta_{pq} & 
\{f\ssc{p}, f_q\} &= \{f_p^\dagger, f_q^\dagger\} = 0.
\end{align}
Because these relations discriminate between particles and antiparticles, we
call them \emph{Dirac} fermions.

One can always mathematically split a Dirac fermion into a pair of \emph{Majorana}
fermions. The corresponding elementary Majorana fermion operators are
\begin{align}
\label{eq:Dirac-to-Majorana}
c\ssc{2p} &:= f_{p}^\dagger + f\ssc{p} &
c\ssc{2p+1} &:= i(f_{p}^\dagger - f\ssc{p}).
\end{align}

The induced Majorana fermion relations, which do \emph{not} discriminate
between particles and antiparticles, are
\begin{align}
\label{eq:Majorana-commutation-relations}
\{c_{p}, c_q\} &= 2\delta_{pq} & 
c_p^\dagger &= c\ssc{p}.
\end{align}

Convenient derived fermionic operators include the \emph{mode number
operator}, 
\begin{align}
\label{eq:mode-number-operator}
{n}\ssc{p} :=&\ f_p^\dagger f\ssc{p}
 = \frac{1}{2}\left(1 + ic\ssc{2p} c\ssc{2p+1}\right),
\end{align}
and the \emph{total mode number operator},
\begin{align}
{n} := \sum_p {n}_p.
\end{align}

Fermionic Hilbert space, an example of a \emph{Fock space}, is the
completion of the infinite direct sum of antisymmetrized eigenspaces of
${n}$.  Its standard basis is the set of \emph{Fock states}, which are
states having definite eigenvalues for all ${n}_p$.  The eigenvalues $N_p$
for $n_p$ are restricted to be $0$ or $1$ by the commutation relations,
while the eigenvalue $N$ for $n$ can take on any non-negative integer
value, or even be countably infinite.

Fermionic operators act on Fock state modes in a way that depends on the
occupations of all previous modes, relative to some prescribed ordering of them: 
\begin{align}
\label{eq:annihilation-op}
f_p
|\ldots, N_{p}, \ldots \rangle
=&
(-1)^{\sum_{k = 0}^{p-1} N_k} 
\,N_p|\ldots, N_{p} - 1, \ldots \rangle \\
\label{eq:creation-op}
f_p^\dagger
|\ldots, N_{p}, \ldots \rangle
=&
(-1)^{\sum_{k = 0}^{p-1} N_k}(1 - N_p)
|\ldots, N_{p} + 1, \ldots \rangle.
\end{align}

A \emph{local} fermionic Hamiltonian is one that can be expressed as a sum of
terms, each of which acts on Fock states in a way that depends only a constant
number of fermionic modes.  In order to avoid a dependence on the prescribed
ordering used to label the fermion modes, local Hamiltonians have the feature
that they only contain an even number of fermionic operators. 

A final fermionic operator worth noting is the \emph{total fermionic parity
operator},
\begin{align}
Q := (-1)^{{n}}.
\end{align}
Because each term in a local Hamiltonian $H$ acts on an even number of modes, it
necessarily obeys $[H, Q] = 0$, conserving total fermionic parity.  Such
a Hamiltonian might or might not also obey $[H, {n}] = 0$.  If it does not obey this,
then the total mode number is not conserved.  This is the case for Hamiltonians
that include BCS-like interactions \cite{Bardeen:1957a, Bardeen:1957b} of the
form
\begin{align}
\sum_{p,q} (f\ssc{p} f\ssc{q} + f_q^\dagger f_p^\dagger).
\end{align}

In terms of Majorana fermion operators, the most general form of an $N$-mode local
fermionic Hamiltonian with only one-body and two-body interactions is
\begin{align}
H &= \sum_{p, q = 0}^{2N - 1} h_{pq} c_p c_{q} 
    + \sum_{p, q, r, s = 0}^{2N - 1} g_{pqrs} c_p c_q c_r c_s,
\end{align}
where the $h_{pq}$ and $g_{pqrs}$ are real coefficients.
%
%Like the Dirac fermionic Hamiltonian above, it contains only terms acting on an
%even number of modes, and thus similarly will conserve total fermionic parity.
%Additionally, note that $[c_p c_q, c_r c_s] = 0$ if $p \ne q \ne r \ne s$; while
%single Majorana fermion operators on distinct modes anticommute,
%\emph{quadratic} Majorana fermion operators on distinct modes \emph{commute}.
%
This is the class of fermionic Hamiltonians we will show how to simulate
fault-tolerantly.

%%%%%%%%%%%%%%%%%%%%%%%%%%%%%%%%%%%%%%%%%%%%%%%%%%%%%%%%%%%%%%%%%%%%%
% Subsection
%
\subsection{The Jordan-Wigner transformation}
\label{sec:jordan-wigner}

As noted in Sec.~\ref{sec:introduction}, mapping fermions to qubits is a task
that necessarily incurs a non-negligible overhead.  The reason for this
essentially boils down to the fact that the commutation algebra of qubit Pauli
operators is much different than that of fermions. Pauli operators can
anticommute on the same qubit, but they always commute on different qubits.  On
the other hand, fermion creation and annihilation operators anticommute even
when they act on different modes.

While there are many fermion-to-qubit mappings, as noted in
Sec.~\ref{sec:introduction}, two general paradigms can be considered.
When $n$ fermions are represented with $n$ qubits, as in \cite{Jordan:1928a,
Bravyi:2000a, Seeley:2012a, Jiang:2020a}, annihilation and creation operators cannot
be asymptotically local; on average, each fermionic operator must be mapped to a
qubit operator acting on at least $\log_3(2n)$ qubits, if the fermionic algebra
is to be represented faithfully \cite{Jiang:2020a}. Alternatively, for lattices
and other systems with limited connectivity, that structure can be exploited to
allow constant-weight fermionic operators to be mapped to constant-weight fermionic
operators. However, this requires a number of additional qubits that scales with the
connectivity of the target system \cite{Verstraete:2005a, Havlicek:2017a, Jiang:2018a,
Steudtner:2019a, Setia:2019a, Derby:2020a, Kirby:2021a, Chen:2022a}. Hybrids between
these approaches are also possible \cite{OBrien:2022a}, with some of the advantages of
each but still dependent on limited connectivity to be efficient.

While the ternary tree mapping \cite{Jiang:2020a} nearly saturates the 
$\log_3(2n)$ bound for a general-purpose mapping, we describe the asymptotically
less efficient, but simpler, Jordan-Wigner transformation \cite{Jordan:1928a}
here instead.  Our reason for doing this is that, at the logical Majorana
fermion level, the reduced-stack architecture we explore shares many
similarities with this transformation.

In the Jordan-Wigner transformation, the fermion creation and annihilation
operators for mode $p$ in an $N$-fermion system are represented by weight-$p$
Pauli operators on qubits $0, \ldots, N-1$ as follows:

\begin{align}
\label{eq:fermion-jordan-wigner-transformation}
f_p &=
 \frac{1}{2} (Z_0 \otimes \cdots \otimes Z_{p-1})
 \otimes (X_{p} + iY_{p}) \\
f_p^{\dagger} &=
 \frac{1}{2} (Z_0^{\vphantom{\dagger}} \otimes \cdots \otimes
 Z_{p-1}^{\vphantom{\dagger}})
 \otimes (X_{p}^{\vphantom{\dagger}} - iY_{p}^{\vphantom{\dagger}}).
\end{align}

Somewhat more simply, the corresponding Majorana fermion operators are mapped as
follows under the Jordan-Wigner transformation:
\begin{align}
\label{eq:Majorana-fermion-jordan-wigner-transformation}
c_{2p} &=
 (Z_0 \otimes \cdots \otimes Z_{p-1})
 \otimes X_{p} \\
c_{2p+1} &=
 (Z_0 \otimes \cdots \otimes Z_{p-1})
 \otimes Y_{p}.
\end{align}
Inverting this mapping yields
\begin{align}
\label{eq:JW-inverted-X}
X_{p} &= (-i)^p c_0 \ldots c_{2p} \\
\label{eq:JW-inverted-Z}
Z_p &= -ic_{2p}\,c_{2p+1}.
\end{align}
Notably, the operators that map to single-site Pauli $X$ operators are
odd-weight Majorana operators, which do not preserve $Q$, the global fermionic
parity.

%%%%%%%%%%%%%%%%%%%%%%%%%%%%%%%%%%%%%%%%%%%%%%%%%%%%%%%%%%%%%%%%%%%%%
% Subsection
%
\subsection{Majorana fermionic quantum circuits}
\label{sec:Majorana-fermionic-quantum-circuits}

In the standard Majorana fermionic quantum circuit (MFQC) model of quantum
computation \cite{Bravyi:2000a}, solving a computational problem is a
three-step process:
\begin{enumerate}

\item Use a classical computer to select a quantum circuit from
a $\mathsf{P}$-uniform family of Majorana fermionic quantum circuits \cite{Nishimura:2002a}.

\item Execute the MFQC on a quantum computer.

\item Return the classical result, say, as a bit string.

\end{enumerate}

In this process, each MFQC in the family is expressible as a sequence of
elementary operations on a collection of Majorana fermions.  The elementary
operations include preparations, measurements, and quantum coherent
operations.  Here, as is common, we restrict our attention to circuits in
which each elementary operation acts on only a constant number of operands.
Generally, to ensure that elementary Majorana operations can be
generated by local Hamiltonians, they are restricted to act on only even numbers
of Majorana fermions, which ensures global fermionic parity preservation. 
However, we will only consider elementary operations that act on two or four
Majorana fermions at a time.
If the set of elementary operations can be used to approximate any
parity-preserving transformation on the Majorana fermions arbitrarily well by
a sufficiently long circuit, then the set of operations is said to comprise
a universal gate set.  (We use the term ``gate set'' even when the set contains
not just coherent gates but also preparation and measurement operations.)

In their landmark paper defining the standard MFQC model \cite{Bravyi:2000a},
Bravyi and Kitaev presented several universal gate sets.  The one we consider
realizing here with logical Majorana fermions is one of these, adapted by Li in
Ref.~\cite{Li:2018a} so that it uses only preparations, measurements, and
Majorana exchanges on a collection of Majorana fermions indexed by the variables
$p$, $q$, $r$, and $s$.  The fact that exchanging Majoranas can implement
nontrivial gates is related to the fact that Majoranas are a model of Ising
anyons, which are particles that interact topologically in (2+1) spacetime
dimensions.  For more details on the connection to topological quantum
computing, including the connection between Majorana fermions and Ising anyons,
see, for example, the textbooks by Wang \cite{Wang:2010a} and Pachos
\cite{Pachos:2012a}.
\begin{enumerate}

\item Prepare a $+1$ eigenstate of the \emph{Majorana fusion} operator,
sometimes also called the \emph{Majorana exchange} operator,
$F_{pq} = ic_p c_q$.  This operator is closely related to, but not the same as,
the mode number operator of Eq.~\eqref{eq:mode-number-operator}.

\item Measure the two-fermion observable $F_{pq}$ (either destructively or
non-destructively.\footnote{By ``non-destructive,'' we mean that the
post-measured quantum state is an eigenstate of the observable measured, as per the
standard von Neumann prescription \cite{Neumann:1955a}.  In a destructive
measurement, only the classical measurement outcome remains.}).

\item Measure the four-fermion observable
$F_{pq}F_{rs}$
nondestructively.

\item Prepare the ``magic state'' $|T\>_{pqrs}$, which is the $+1$
eigenstate of the following observables:
\begin{gather}
\label{eq:T-state-MF1}
F_{pq} F_{rs}
\\[1ex]
\label{eq:T-state-MF2}
\frac{F_{pq} + F_{pr}}
{\sqrt{2}}.
\end{gather}

\item Apply the (unitary) Majorana exchange gate $F_{pq}$.

\end{enumerate}

\begin{figure}[H]
\begin{center}
  \includegraphics[width=0.95\columnwidth]{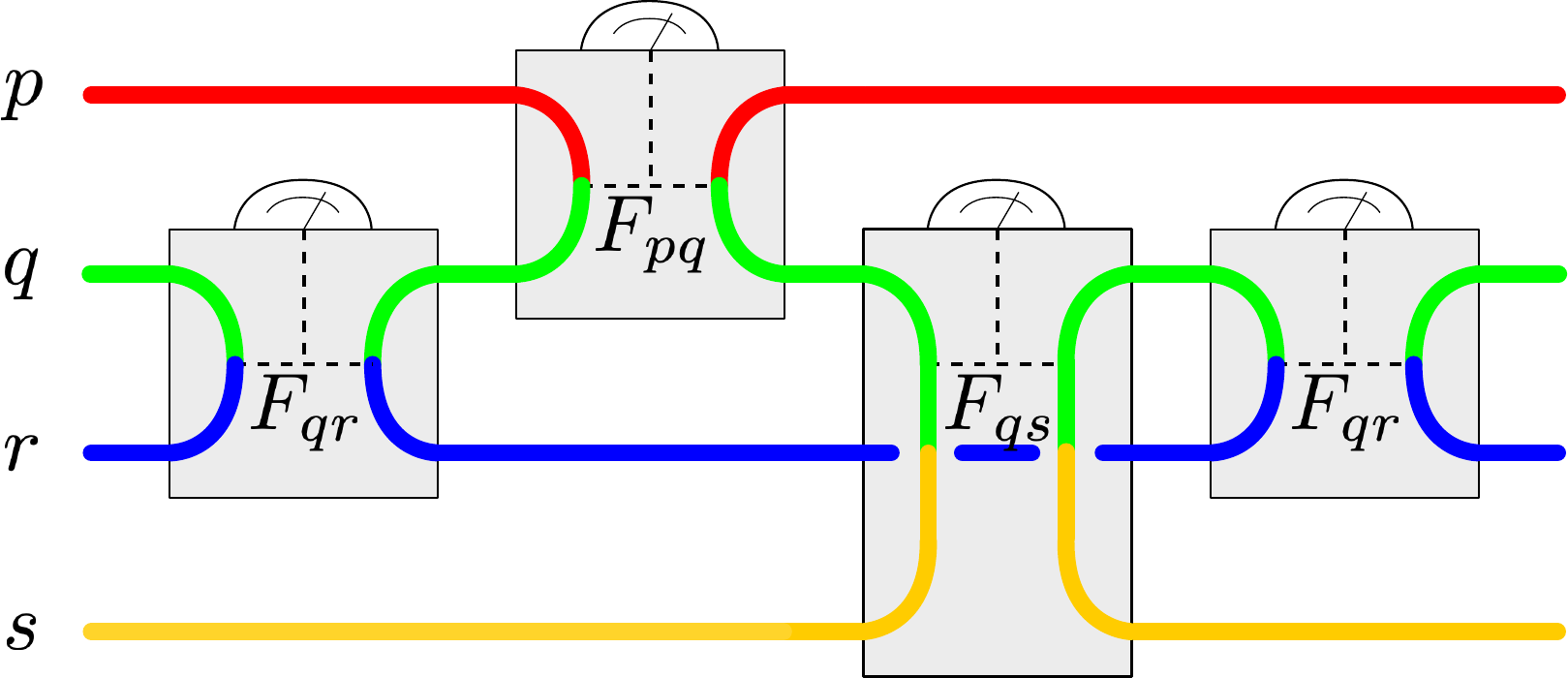}
\caption{\small{\label{fig:ccw-swap-circuit}(Color online.) A fermionic quantum
circuit for realizing the coherent counterclockwise exchange $F_{ps}$ of
Majorana fermions using only non-destructive measurements on pairs of Majorana
fermions.  The information is exchanged between the Majorana fermion modes
labeled $p$, $q$, $r$, and $s$, even though the carriers themselves are not
exchanged (indicated by colored worldlines moving from left to right).
Generally, the measurement of the fusion operators can yield non-trivial
outcomes, which can be handled by adaptive logic, such as with a ``forced
measurement'' protocol \cite{Bonderson:2008a} or ``Majorana frame tracking''
\cite{Zheng:2016a}; the circuit depicted here imagines that all fusion
measurement outcomes are trivial.}}
\end{center}
\end{figure}

By using Majorana exchange frame tracking \cite{Zheng:2016a}, reminiscent of
Clifford frame tracking for qubits \cite{Chamberland:2018a, Litinski:2019a}, one
can dispense with the final coherent gate in the Li gate set, converting the
gate set into one comprised solely of measurement and preparation operations.
This basis allows one to realize measurement-based universal topological quantum
computing~\cite{Bonderson:2008a, Barkeshli:2016a, Zheng:2016a, Tran:2020a} in
a fashion reminiscent of fusion-based quantum computing \cite{Bartolucci:2021a}.
Fig.~\ref{fig:ccw-swap-circuit} depicts an example of a Majorana fermion circuit
that implements the coherent Majorana exchange gate $F_{ps}$ using only
non-destructive two-fermion measurements.

%%%%%%%%%%%%%%%%%%%%%%%%%%%%%%%%%%%%%%%%%%%%%%%%%%%%%%%%%%%%%%%%%%%%%
% Subsection
%
\subsection{Majorana fermion stabilizer codes}
\label{sec:Majorana-fermion-stabilizer-codes}

A \emph{Majorana fermion stabilizer code} \cite{Bravyi:2010b}, or Majorana
stabilizer code for brevity, is the simultaneous $+1$ eigenspace of a
collection of commuting, Hermitian, even-weight Majorana operators.  The
evenness constraint ensures that these operators are fermion-parity
preserving, and hence physically observable.  Following the language used
for qubit stabilizer codes \cite{Gottesman:1997a}, we say these operators
generate the code's \emph{stabilizer group}, and each operator is called a
\emph{stabilizer generator}, or sometimes just a \emph{stabilizer} or a
\emph{check} for brevity, because they ``stabilize'' the codespace and are
what are measured to ``check'' for errors.  More generally, in a
\emph{subsystem} Majorana stabilizer code, the measured checks need not
commute and the stabilizer group is defined to be the center of the
\emph{check group}.  Whether for subspace or subsystem Majorana
stabilizer codes, the \emph{logical group} is the check group's normalizer.

Without loss of generality, each check in a Majorana stabilizer code can be
written as $S_{\sigma}$, where $\sigma$ indicates the set of modes
$V_{\sigma}$ on which it has support.  Each logical operator $L$ is
supported on a set of modes $V_L$ obeying $|V_L \cap V_{\sigma}| \equiv 0 \bmod
2$ for each ${\sigma}$ to ensure that the logical group and stabilizer group
commute.  Mathematically, each check and logical operator can be expressed as
\begin{align}
\label{eq:stabilizer-majorana}
S_{\sigma} &:= i^{|V_{\sigma}|/2} \prod_{q \in V_{\sigma}} c_q \\
\label{eq:logical-majorana}
L &:= \eta_L \prod_{q \in V_L} c_q,
\end{align}
where $\eta_L \in \{\pm 1, \pm i\}$ is a phase.

The \emph{distance} of a Majorana fermion stabilizer code is the minimum
nonzero weight of its logical group's elements.
Unlike qubit codes, a Majorana fermion stabilizer code of distance $d$
cannot necessarily correct all Majorana errors of weight $\lfloor (d - 1)/2
\rfloor$ or less, because the codes treat even and odd logical operators on the same
footing, exposing them, \eg, to ``quasiparticle poisoning'' errors from
weight-one Majorana fermion operators, which do not conserve fermionic
parity locally, but might do so when the environmental degrees of freedom of
the bath are taken into consideration \cite{Vijay:2017a}.  With clever
concatenation techniques, this can be avoided, as will be discussed later.

A \emph{fermion stabilizer code} is a Majorana stabilizer code on an even
number ($2n)$ of Majorana fermions \cite{Bravyi:2010b, Vijay:2017a}.  One
can show that a fermion stabilizer code's logical group is isomorphic to the
Pauli group on $k = n - |\{S_\sigma\}|$ logical qubits, so that one can think
of such codes as encoding $k$ logical qubits in $2n$ physical Majorana
fermions \cite{Bravyi:2010b}.  Following Ref.~\cite{Vijay:2017a}, we use
$[\![n, k, d]\!]_f$, or alternatively $[\![2n, k, d]\!]_m$, to denote a
fermion stabilizer code that encodes $k$ qubits to distance $d$ in $n$ Dirac
fermions, or equivalently, in $2n$ Majorana fermions.

A \emph{Majorana surface code} is a Majorana stabilizer code defined by an
embedding of a graph into a surface.  Generally, one can use a rotation
system to define such codes \cite{Sarkar:2021a}.  Here, we only need to
consider a subclass of Majorana surface codes, first described by Litinski
and van Oppen \cite{Litinski:2018a}, that are defined by
face-three-colorable graphs embedded on a disk in which one associates
Majorana fermions with vertices and checks with faces.  (The ``face''
outside the graph in the disk might not be able to be consistently colored
with all the other faces \cite{Sarkar:2021a}.) 

%%%%%%%%%%%%%%%%%%%%%%%%%%%%%%%%%%%%%%%%%%%%%%%%%%%%%%%%%%%%%%%%%%%%%
% Subsection
%
\subsection{Majorana cycle codes}
\label{sec:Majorana-cycle-codes}

A \emph{Majorana cycle code} is a $[\![n+1, n, 2]\!]_f$ Majorana surface code
defined by a $(2n+2)$-vertex cycle graph embedded on a disk.  Each Majorana
cycle code has a single check, corresponding to the product of all of the
Majorana fermion operators on the vertices, up to a phase consistent with
Eq.~\eqref{eq:stabilizer-majorana}.  The constraint that the codes are $+1$
eigenstates of this check operator is just the even-parity constraint required
of any collection of indistinguishable fermions.

There are many ways of choosing logical Pauli operator bases for these codes.
The simplest nontrivial Majorana cycle codes are called the \emph{tetron code}
and the \emph{hexon code}, encoded into four and six Majorana fermions
respectively~\cite{Bravyi:2006a, Hastings:2017a, Litinski:2018a}.  For the
tetron code, we choose the logical operator basis to be
\begin{align}
\label{eq:xbar-zbar-tetron}
\overline{X} &= -ic_0 c_1 
&
\overline{Z} &= -ic_1 c_2.
\end{align}
We include overall minus signs in the $\eta_L$ phases used in these definitions
to make the relationship between this encoding and the inverse Jordan-Wigner
transformation in Eqs.~(\ref{eq:JW-inverted-X})--(\ref{eq:JW-inverted-Z}) clearer.
The stabilizer group for this code is generated by a single element, which from
Eq.~\eqref{eq:stabilizer-majorana} is
\begin{align}
S &= -c_0 c_1 c_2 c_3.
\end{align}

\begin{figure}[H]
\center{
  \subfigure[\
Tetron.]{\includegraphics[height=1.0in]{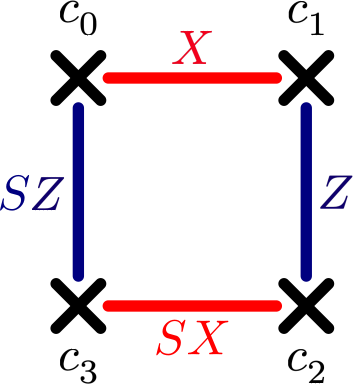}}
\qquad
  \subfigure[\label{fig:hexon-code}\
Hexon.]{\includegraphics[height=1.0in]{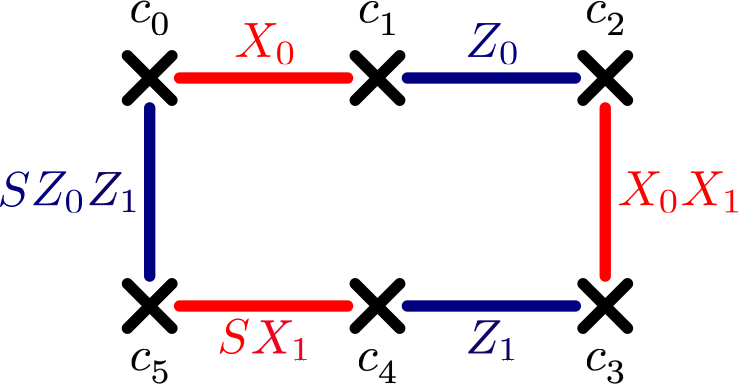}}
}
\caption{\small{\label{fig:tetron-hexon}(Color online.) Logical operator bases
for some simple Majorana cycle codes.}}
\end{figure}

For the hexon code, we choose the logical operator basis to be
\begin{align}
\overline{X}_0 &= -ic_0 c_1
&
\overline{Z}_0 &= -ic_1 c_2  \\
\overline{X}_1 &= -c_0 c_1 c_2 c_3
&
\overline{Z}_1 &= -ic_3 c_4,
\end{align}
with a stabilizer group generated by the operator
\begin{align}
S &= -ic_0 c_1 c_2 c_3 c_4 c_5.
\end{align}
These choices are captured in Fig.~\ref{fig:tetron-hexon}.

\begin{figure}[H]
\center{
  \includegraphics[height=1.0in]{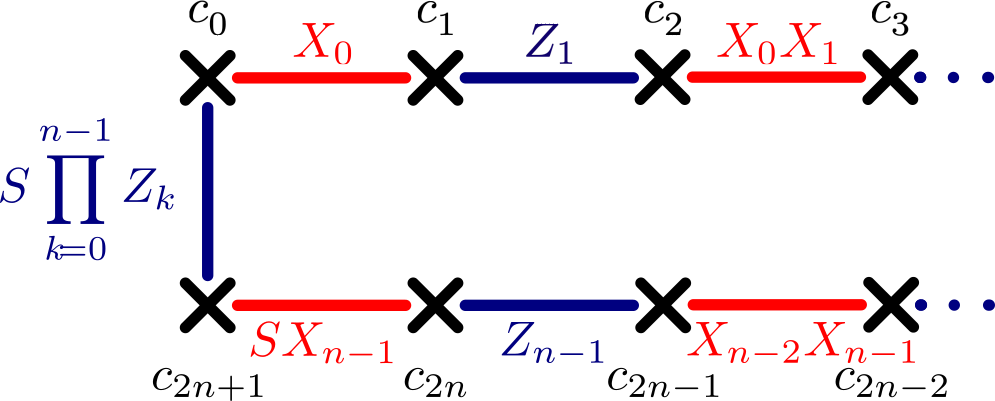}
}
\caption{\label{fig:two-n-plus-two-on}(Color online.) Logical operator basis for
a general Majorana cycle code.}
\end{figure}

Generally, we choose the logical operator basis for a Majorana cycle code to be
\begin{align}
\label{eq:xbar-zbar-ennon}
\overline{X}_p &= (-i)^{p+1} (c_0\ldots c_{2p+1})
\\
\overline{Z}_p &= -ic_{2p+1} c_{2p+2}.
\end{align}
This choice is depicted diagrammtically in Fig.~\ref{fig:two-n-plus-two-on}.

The Majorana cycle code is a Jordan-Wigner-like code; it differs from the
inverted Jordan-Wigner transformation described in
Eqs.~(\ref{eq:JW-inverted-X})--(\ref{eq:JW-inverted-Z}) by the addtion of two
extra Majorana fermions.  These serve to ensure that every $\overline{X}_p$
preserves global fermionic parity: each logical $\overline{X}_p$ now acts on one
more Majorana mode, giving it even weight, and each of the logical $\overline{Z}_p$
operators is defined on a pair of Majorana modes that are shifted by one relative to
the inverse Jordan-Wigner transformation to accommodate this.  Notice that the
logical operators defined on the mode $c_{2n+1}$ are elusive.  The $X$-like and
$Z$-like logical operators defined on them are equivalently expressed as
a product of all of the other $X$-like and $Z$-like logical operators,
respectively, up to multiplication by the lone stabilizer generator.

A large Majorana cycle block code like the one depicted in
Fig.~\ref{fig:two-n-plus-two-on} is nearly twice as efficient at encoding qubits
as a collection of tetron block codes combined via a tensor product, which yield
effectively only a $[\![2n, n, 2]\!]_f$ code.  For this reason, and to simplify
our constructions later, we only consider encodings into large Majorana cycle
block codes \cite{LogicalMajoranaPatchEndnote}. We discuss
how to move between these encodings in Appendix~\ref{sec:code-deformation}.

%%%%%%%%%%%%%%%%%%%%%%%%%%%%%%%%%%%%%%%%%%%%%%%%%%%%%%%%%%%%%%%%%%%%%
% Subsection
%
\subsection{Logical Majorana codes}
\label{sec:qubit-surface-code-patches}

Like Majorana surface codes, qubit surface codes \cite{Kitaev:1996a,
Kitaev:1997a, Kitaev:1997b, Bravyi:1998a, Freedman:2001b, Bombin:2006a,
Bombin:2007d, Anderson:2011a} can be defined using a rotation system that
describes a graph embedding combinatorially \cite{Sarkar:2021a}.  As with
Majorana surface codes, we will only consider a narrow subclass of such codes,
namely those defined by face-two-colorable (``checkerboardable'') graphs
containing squares and digons embedded on a disk in which one associates qubits
with vertices and checks with faces.  (The ``face'' outside the graph in the
disk might not be able to be consistently colored with all the other faces \cite{Sarkar:2021a}.)
A check of one face color can be associated with a tensor product of Pauli $X$
operators on its incident qubits; a check of the other face color can be
associated with a tensor product of Pauli $Z$ operators on its incident qubits.
Alternatively, all checks can be given the same local structure by a set of
local basis changes, turning it into a so-called $XZZX$ code
\cite{Bonilla:2020a}.  Because these qubit surface codes are embedded on a disk,
we call them qubit surface-code patches. 

\begin{figure}[H]
\center{
  \subfigure[\ Tetron.]{\includegraphics[height=1.0in]{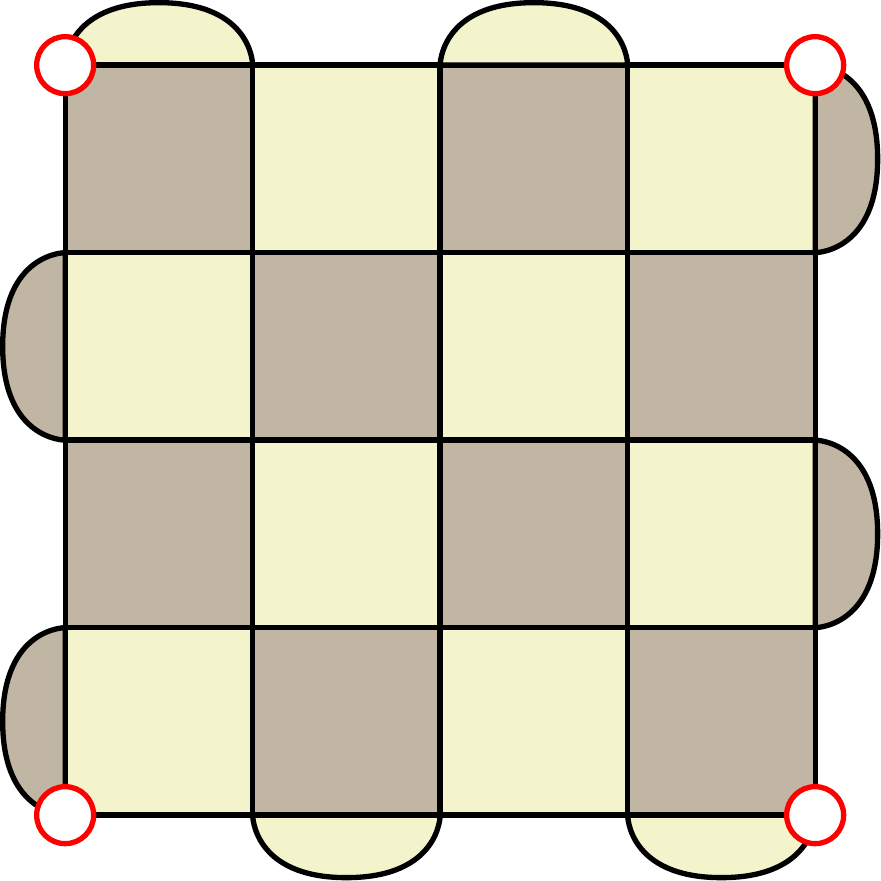}}
\qquad
  \subfigure[\ Hexon.]{\includegraphics[height=1.0in]{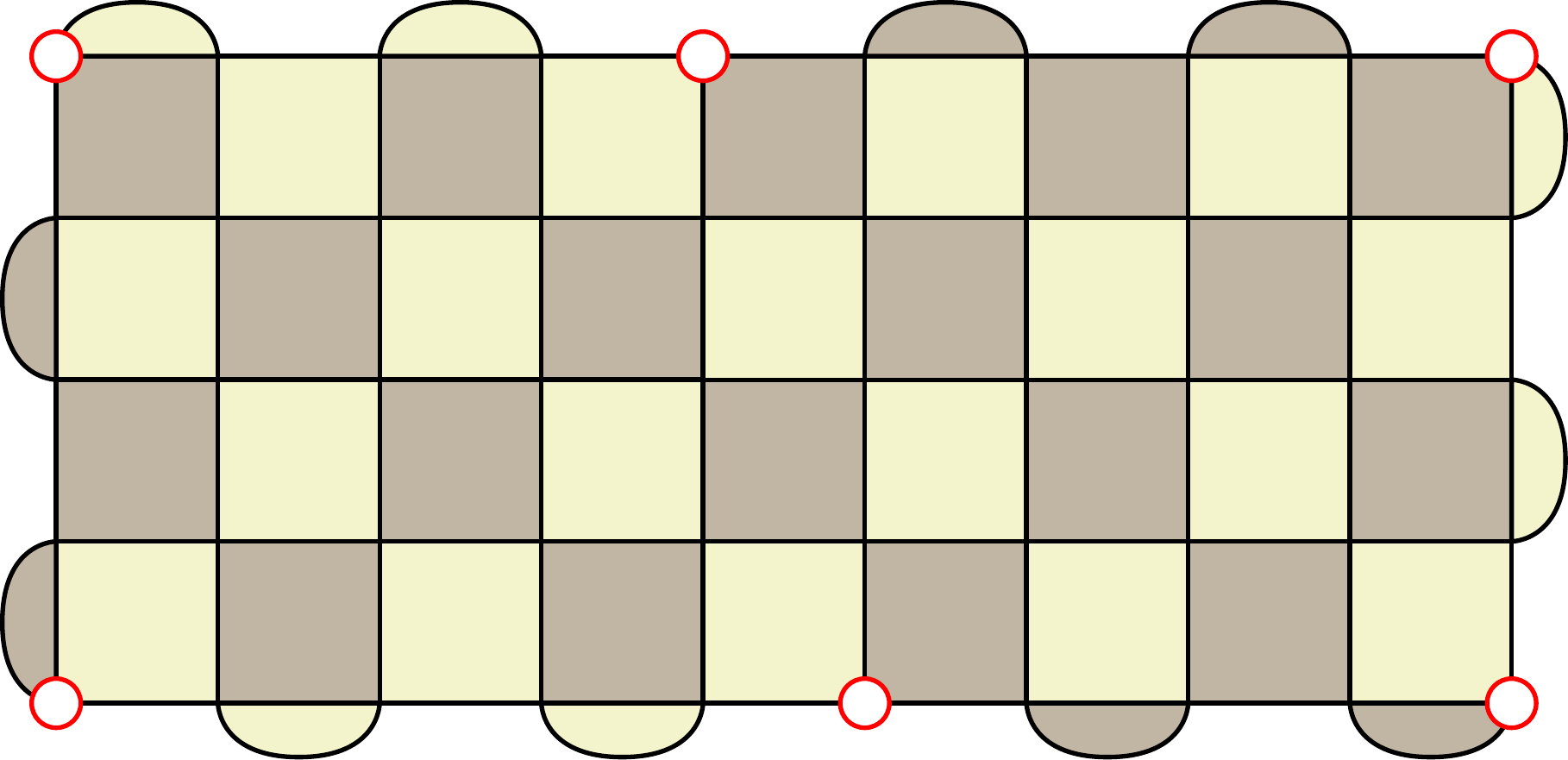}}
}
\caption{\label{fig:tetron-hexon-surface-codes}(Color online.) Distance-five
versions of surface code patches that realize logical Majorana fermions as
``corner twist defects,'' depicted by white circles with red boundaries.}
\end{figure}

The number of logical qubits encoded by a surface code patch is related to its
number of boundary components, where each boundary component can be associated
with a logical operator, as described in Ref.~\cite{Kesselring:2018a}.  Because
of this, the perimeter can be described by a cycle graph, in exactly the same
way that Majorana cycle codes are described.  The locations where the boundary
type changes are called \emph{corner twists}, following terminology used in
Refs.~\cite{Bombin:2010b, Kitaev:2012a, You:2012a, You:2013a, Yoder:2017a,
Brown:2017a, Kesselring:2018a, Brown:2020a, Lavasani:2018a, Barkeshli:2019a,
Zhu:2020a, Bombin:2021a}.  Examples of surface-code patches with perimeters
that can be described by the $C_4$ (tetron) and $C_6$ (hexon) cycle graphs
are depicted in Fig.~\ref{fig:tetron-hexon-surface-codes}.

The similarity between the codes in Fig.~\ref{fig:tetron-hexon} and
Fig.~\ref{fig:tetron-hexon-surface-codes} is not mere coincidence.  The corner
twist defects act like logical Majorana fermions (or, more precisely, like
$\ZZ_2$-crossed braided tensor categories \cite{Barkeshli:2019a}), as pointed
out in Refs.~\cite{Bombin:2010b, Brown:2017a, Kesselring:2018a}.  In other
words, these codes essentially encode logical Majorana fermions in physical
qubits.  To disambiguate from the term ``Majorana fermion code,'' which
describes a code that works the other way around by encoding physical Majorana
fermions into logical qubits \cite{Kitaev:1997a, Bravyi:2010b, Vijay:2015a,
Vijay:2016a, Vijay:2017a, Hastings:2017a, Litinski:2017a, Li:2018a,
Litinski:2018a}, we will call these \emph{logical} Majorana fermion codes, or
just logical Majorana codes, for short.

Upon closer examination, one can see that the logical Majorana operators for
the tetron and hexon qubit surface codes are themselves further encoded in
the tetron and hexon codes, respectively.  These qubit surface-code patches
are therefore \emph{concatenated} codes, in which physical qubits realize
logical Majorana fermions, which in turn realize logical qubits.  A helpful
consequence is that quasiparticle poisoning errors are not a concern,
because errors on physical qubit-level operations, once decoded with the
lower-level logical Majorana code, translate at worst to even-weight logical
Majorana operators in the upper-level tetron or hexon code.  This means that
a distance-$d$ qubit tetron or hexon surface code protects its encoded
logical Majorana fermions against all physical errors of weight
$\lfloor(d-1)/2\rfloor$ or less.  These properties hold for the straightforward
generalization to general qubit surface-code patches that are associated with
Majorana cycle codes.

%%%%%%%%%%%%%%%%%%%%%%%%%%%%%%%%%%%%%%%%%%%%%%%%%%%%%%%%%%%%%%%%%%%%%
% Subsection
%
\subsection{Fault-tolerant surface-code computation with logical Pauli
measurements and preparations}
\label{sec:ft-sc}

In the ordinary qubit quantum circuit model, there is a universal gate basis
that consists solely of one- and two-qubit preparations and measurements on
pairs of qubits (labeled $p$ and $q$) that is similar to the Li gate basis
described in Sec.~\ref{sec:Majorana-fermionic-quantum-circuits}.  The gate basis
is as follows:
\begin{enumerate}

\item Prepare a $+1$ eigenstate of the Pauli operators $X_p$ and $Z_p$.

\item Measure the Pauli operators $X_p$ and $Z_p$ (either destructively or
non-destructively).

\item Measure the Pauli operators $X_p X_q$ and $Z_p Z_q$
nondestructively.

\item Prepare the ``magic state'' $|T\>_p$, which is the $+1$
eigenstate of the following observable:
\begin{gather}
\frac{X_{p} + Z_{p}}
{\sqrt{2}}.
\end{gather}

\end{enumerate}

Although logical Pauli operators can be represented by simple strings of
physical Pauli operators between twists in the surface code, measuring and
preparing these operators fault-tolerantly is not as simple as just measuring or
preparing the qubits along the relevant strings, because the individual qubit
operations might be faulty.  Nevertheless, several fault-tolerant protocols for
realizing these exist.  Here, we review one preparation protocol
\cite{Landahl:2014a} and three measurement protocols: one based on interior
twist braiding~\cite{Brown:2017a}, one based on corner twist lattice
surgery~\cite{Litinski:2019a}, and one based on twist teleportation via ``portal
pairs''~\cite{Bombin:2021a}.

A helpful primitive for all of these protocols is a fault-tolerant procedure for
bringing corner twists to the interior (and vice versa) of surface code patches;
one way of doing this is described in detail in Ref.~\cite{Brown:2017a}.
Interior twists come in one of two varieties.  An interior twist can either be
a disclination (rotational) twist or a dislocation (translational) twist.
Either type causes the colorability of plaquettes in the surface-code lattice to
be frustrated; the latter have a convenient representation that does not disturb
the locations of vertices in the lattice, so we will focus our attention on this
type for pedagogical clarity.

\begin{figure}[H]
\center{
  \includegraphics[height=2.0in]{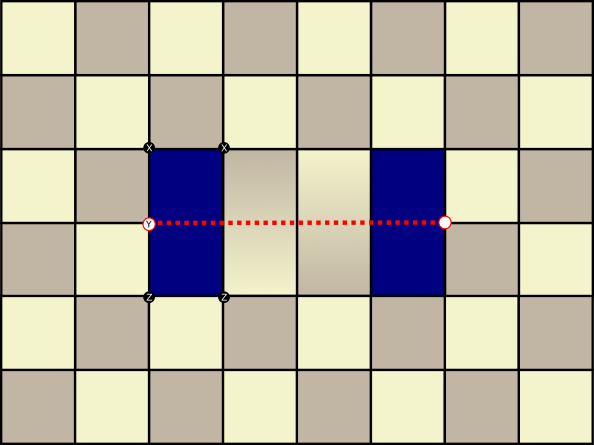}
}
\caption{\label{fig:xxyzz-defect}(Color online.) A logical qubit implemented as
a pair of dislocation twist defects, represented by ``missing'' $XXYZZ$ checks.}
\end{figure}

A small interior dislocation twist can be represented by a weight-five $XXYZZ$
check that spans the space that two ordinary weight-four $XXXX$ and $ZZZZ$
checks would have occupied.  The code distance of the associated logical qubit
represented by this twist is the minimum of the perimeter of the twist (in this
case, five) and the distance of the twist to the nearest other twist.
Manifestly, then, a code with just a single twist cannot encode a logical qubit.
The code distance can be increased by removing adjacent checks to the twist and
separating it farther away from other twists.  Pairs of twists can be connected
by ``defect lines,'' which connect the qubits in each twist on which the $Y$
operator has support along a collection of qubits that have been removed from
the lattice to accomodate the twist; Fig.~\ref{fig:xxyzz-defect} depicts an
example of two weight-five interior twists whose separation from each other is
just three qubits apart.

Most fault-tolerant preparation protocols for logical qubits in the surface code
focus on preparing a single logical qubit in the tetron surface code.  This is
because, in principle, preparations for isolated tetron surface codes can be
combined by first bringing the corner twists into the interior as puncture
defects, and then stitching them together via lattice surgery, as described, for
example, in Refs.~\cite{Horsman:2012a, Landahl:2014a, Cesare:2014a}.  The
preparation protocol for a single tetron is fairly simple: one prepares all
physical qubits in the eigenstate of the desired Pauli operator, $X$ or $Z$, and
then performs fault-tolerant syndrome extraction on the qubits.  One way to
prepare a low-fidelity version of the magic state (to ``inject it'' into the
code), as described in Ref.~\cite{Landahl:2014a}, is to first prepare four
mini-patches, one in each quadrant, surrounding a physical qubit prepared in the
$|T\>$ state.  Then one measures the checks three times and performs
fault-tolerant decoding on the result.  Although this only suppresses errors to
second-order in the errors of the physical operations, the preparation of the
physical $|T\>$ state itself is only good to first-order in this error.  The
resultant low-fidelity state can be distilled to any desired precision (limited
by the code distance) by any of a number of magic-state distillation protocols,
such as the Bravyi-Kitaev protocol \cite{Bravyi:2005a}, using the other
fault-tolerant operations in the universal set.

Ref.~\cite{Brown:2017a} describes a protocol for how to measure a logical qubit
represented by a pair of interior twists.  In the protocol, one entangles the
twist pair with an auxiliary logical qubit represented by a pair of (non-twist)
punctures (\viz, missing connected check regions). The protocol requires
braiding one of the punctures around the twist pair by a sequence of internal
code deformations, followed by destructively measuring the auxiliary logical
puncture pair that represents the logical qubit. Indeed, this protocol can be used
to measure the product of any number of logical qubits stored in an even number
of twists, simply by braiding the punctures along the correct path. This protocol
requires a greater separation between twists than the other two alternatives we
describe next, because it is necessary to leave space for the punctures to move
around without compromising the distance of the code. However, it doesn't
require any unusual hardware features beyond the ability to turn particular
stabilizer generators on and off.

By moving all interior twists to corner twists on the boundary, one can measure
an arbitrary logical Pauli product operator using the the lattice-surgery
protocols described in Ref.~\cite{Litinski:2019a}.  However, this occupies
a length of the surface code's perimeter that is proportional to the number of
twists being measured.  One challenge with this protocol is the limitations on
parallelism it imposes if all logical qubits are confined to a single patch,
such as in a large logical Majorana fermion code when thought of as
a logical-qubit code.  In this case, for example, in order to measure
$\Theta(n)$ quadratic Majorana (single-logical-qubit) or quartic Majorana
(two-logical-qubit) operators on $n$ twists in parallel, one needs a surface code patch
with perimeter at least $\Omega(dn)$, where $d$ is the distance of the code.
While this is possible with a narrow rectangular patch with
twists arranged in a $2 \times \frac{n}{2}$ array, achieving it on the
square-lattice arrays of twists that match the geometry of simulation targets
(and thus support the speedups described in this paper) requires a patch with
highly unconventional structure. A notable
advantage, however, is the ease of $|T\>$ state injection at the boundary, allowing
the use of external $T$-factories rather than needing to locate them within the
main computational patch, as described in Ref.~\cite{Litinski:2019a}.

Finally, Ref.~\cite{Bombin:2021a} describes ``portal pairs'' that can be used to
teleport twists to a future time, allowing direct non-destructive measurements
of any twist operator in a number of rounds equal to the code's distance. This
allows the most compact array of twists, as no physical braiding or movement of
twists whatsoever is required.  However, it requires highly nonlocal physical
operations, potentially even in time as well as in space, and as a result is
only suited to very specific hardware models: in its original conception,
a photonic architecture \cite{Bombin:2021b} with a fiber delay line.

%%%%%%%%%%%%%%%%%%%%%%%%%%%%%%%%%%%%%%%%%%%%%%%%%%%%%%%%%%%%%%%%%%%%%
% Subsection
%
\subsection{Trotter-Suzuki approximation}
\label{sec:trotter-suzuki-background}

For simulation applications run on quantum computers, a common target is the
implementation of time-evolution operator
\begin{align}
U = e^{-it H}
\end{align}
corresponding to a physical Hamiltonian $H$. While this operator is unitary, and
thus at least theoretically possible to implement on a quantum computer, it may
be extremely difficult to implement exactly in practice. Accordingly, several
methods have been proposed for approximate implementations of $U$; here we
discuss the Trotter-Suzuki decomposition~\cite{Suzuki:1990}. It requires the
Hamiltonian be written as a sum of Trotter layers $H_L$, where each layer is
a sum of commuting terms that do not necessarily commute with the terms in other
layers. The exponentiated Hamiltonian
\begin{align}
U = e^{-it H} = e^{-it \sum_L H_L}
\end{align}
is then approximated as
\begin{align}
\tilde{U} = \prod_L e^{-it H_L}.
\end{align}

For practicality of implementation, we will further decompose each Trotter layer
\(H_L\) into a sum of sub-layers, grouping the Pauli products or even-weight
Majorana fermionic operators more finely than the commutation relations would require.
Each of these ``Pauli product layers'' or ``Majorana layers'' should consist only of terms
that are supported on disjoint sets of information carriers (qubits or fermions,
respectively), so that the gates within an exponentiated layer can be implemented
in parallel. Because those sub-layers commute with each other, the sub-decompositions are
exact and do not introduce any additional approximation error into the simulated
evolution.  The resulting circuit depth is proportional to the number of Pauli product
layers or Majorana layers. 

In fault-tolerant quantum computing architectures, it is common that the
fault-tolerant realizations of non-Clifford gates are more costly, in terms of
physical resources like physical qubits and time, than fault-tolerant
realizations of Clifford gates~\cite{Fowler:2012a}.  For this reason, it is also
common to synthesize all fault-tolerant quantum circuits so that they depend on
only one, or only a few, types of non-Clifford gates, such as the $T$ gate,
which is the fourth-root of the Pauli $Z$ gate.  A proxy for the complexity of
fault-tolerant circuits is then the \emph{$T$ count}, which counts the number of
$T$ gates in a quantum circuit.

The $T$ count is unaffected by the choice of Trotter layering, although related
measures, such as the \emph{$T$ depth}, which is the circuit depth of $T$ gates
utilized, may be affected by the choice of Trotter layering.  While $T$ gates
might require many gates to be synthesized fault-tolerantly, say through one or
more ``magic state factories,'' the total runtime of the quantum algorithm is
not necessarily impacted by this cost because, with sufficiently many factories,
``magic'' $|T\>$ states can be supplied with high fidelity on demand whenever a $T$
gate is required \cite{Litinski:2019a}.  Hence, for the purposes of determining
the overall runtime of the fault-tolerant quantum algorithm, one can assign
a depth cost of one to $T$ gates.

Another useful measure of circuit complexity is the overall gate count utilized.
This dictates the size of quantum error correcting code required, because larger
circuits require larger codes to maintain fault tolerance.

%%%%%%%%%%%%%%%%%%%%%%%%%%%%%%%%%%%%%%%%%%%%%%%%%%%%%%%%%%%%%%%%%%%%%
% Subsection
%
\subsection{Block encoding for Hamiltonian simulation}
\label{sec:qubitization-background}

Block-encoding methods~\cite{Low:2016a} have been proposed as an alternative to
Trotter-Suzuki decomposition for simulation of the Fermi-Hubbard model and other
quantum systems~\cite{Babbush:2018a}.

For concreteness, consider a Hamiltonian $H$ that is expressible as a linear
combination of Hermitian unitary operators $H_{\ell}$ with real coefficients $w_{\ell}$
(which, without loss of generality, we take to be positive by absorbing
minus signs into the $H_{\ell}$ terms).  In other words, consider the
Hamiltonian
\begin{align}
H = \sum w_{\ell} H_{\ell}
  \quad \textrm{s.t.}
  \quad w_{\ell} \in \RR,\ w_{\ell} > 0,\ H_{\ell}^2 = \mathds{1}.
\end{align}

From this Hamiltonian, one can construct a pair of oracle circuits.
\textsc{prepare} prepares a superposition $\ket{\mathcal{L}}$ of indices
$\ket{\ell}$ that specify Hamiltonian terms, with coefficients proportional to the
corresponding terms' coefficients $w_\ell$ in the Hamiltonian:
\begin{align}
\label{eq:general-prepare-oracle}
\textrm{\textsc{prepare}}\ket{0} = \sum \sqrt{\frac{w_\ell}{\lambda}} \ket{\ell}
= \ket{\mathcal{L}},
\end{align}
where $\lambda := \sum |w_{\ell}|$.
\textsc{select} applies a Hamiltonian term $H_\ell$ specified by an index
register onto a system register:
\begin{align}
\label{eq:general-select-oracle}
\textrm{\textsc{select}} = \sum_\ell \ket{\ell}\bra{\ell} \otimes H_\ell.
\end{align}

These oracles together satisfy the ``qubitization'' relation
\begin{align}
\left(\bra{\mathcal{L}} \otimes
\mathds{1}\right)\mathrm{\textsc{select}}\left(\ket{\mathcal{L}} \otimes
\mathds{1}\right) = \frac{1}{\lambda} \sum_\ell w_\ell H_\ell = \frac{H}{\lambda},
\end{align}
which means that we can encode the desired spectrum into an associated ``quantum
walk'' operator $\mathcal{W}$ by defining $\mathcal{W}$ as
\begin{align}
\mathcal{W} = \left(2 \ket{\mathcal{L}} \bra{\mathcal{L}} \otimes \mathds{1}
- \mathds{1}\right) \mathrm{\textsc{select}}.
\end{align}

In other words, on the subspace spanned by $|\mathcal{L}\>|k\>$ and the
component of $\mathcal{W}|\mathcal{L}\>|k\>$ that is orthogonal to
$|\mathcal{L}\>|k\>$, the quantum walk operator acts as
\begin{align}
\mathcal{W} = e^{i Y \arccos(E_k/\lambda)},
\end{align}
where $Y$ is the Pauli $Y$ operator and $E_k$ is the $k$-th eigenvalue of $H$.
To find the eigenvalues of $H$, then, it suffices to instead find the
eigenvalues of $\mathcal{W}$.

To perform phase estimation on this operator, we will need to implement a controlled
version of it, which can be done via the following circuit if \textsc{prepare},
\textsc{prepare}$^\dagger$,
and controlled-\textsc{select} are available.

\begin{figure}[H]
\begin{center}
\resizebox{0.45\textwidth}{!}{%
\begin{quantikz}
\lstick{} &[2mm] \qw & \ctrl{1} & \qw & \gate{Z} & \qw & \qw \\
\lstick{${\ket{\ell}}$} &[2mm] \qw\qwbundle{}
& \gate[wires=2]{\mathrm{\textsc{select}}}
& \gate{\mathrm{\textsc{prepare}}^\dagger} & \octrl{-1}
& \gate{\mathrm{\textsc{prepare}}} & \qw \\
\lstick{${\ket{\psi}}$} &[2mm] \qw\qwbundle{} & \qw & \qw & \qw & \qw & \qw \\
\end{quantikz}
}%
\end{center}
\caption{\label{fig:walk-circuit}The controlled qubitized walk operator $\Lambda(\mathcal{W})$.}
\end{figure}
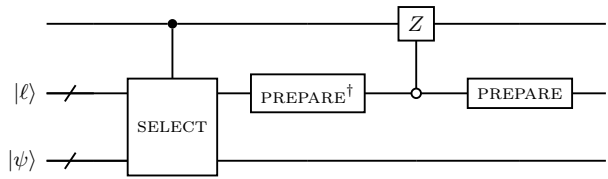
\vspace{0\baselineskip}

The controlled walk operator must be applied $\bigO(\lambda/\Delta E)$ times, where
$\Delta E$ is the desired error bound on the ground state energy.

%%%%%%%%%%%%%%%%%%%%%%%%%%%%%%%%%%%%%%%%%%%%%%%%%%%%%%%%%%%%%%%%%%%%%
% Subsection
%
\subsection{Fermi-Hubbard model system and Hamiltonian}
\label{sec:hubbard-model-background}

In order to demonstrate the techniques we develop herein, we require an exemplar system to
simulate. For the sake of simplicity and ease of application, we will consider the planar
Fermi-Hubbard model for each demonstration.

The Fermi-Hubbard model describes a many-electron system on a lattice; we consider
specifically the case of a two-dimensional square lattice. While originally
constructed to describe correlation effects in the $d$-band of a transition
metal \cite{Hubbard:1963}, under the approximation of a narrow band with
nearest-neighbor and on-site interactions only, it has more recently been
applied to modeling cuprate superconductors \cite{LeBlanc:2015}. The model
Hamiltonian consists of a hopping term and an on-site self-energy term
\begin{align}
\label{eq:hubbard-hamiltonian-dirac}
H_\mathrm{HUB} = -t \sum_{\langle p, q \rangle, \sigma} a^\dagger_{p,\sigma} a\ssc{q,\sigma} + \frac{u}{2} \sum_{p, \alpha \ne \beta} n\ssc{p,\alpha} n\ssc{p,\beta}
\end{align}
where $p$ and $q$ index lattice sites, $\sigma$, $\alpha$, and $\beta$ index spins,
$\langle p, q \rangle$ indicates a sum over pairs of adjacent lattice sites only, and $t$
and $u$ parametrize the interaction strengths. Simulations of the model typically consider
a finite lattice with $N$ sites, before extrapolating results to infinite systems (and the
thermodynamic limit more generally)~\cite{LeBlanc:2015}.

Approximate solutions to the Fermi-Hubbard model---obtained via classical
algorithms including Monte Carlo, embedding methods, density matrix
renormalization group theory, coupled-cluster methods, and multi-reference
Hartree-Fock---show ferromagnetism, superconductivity, and metal-insulator
transitions, depending on Hamiltonian parameters \cite{LeBlanc:2015}. This
combination of varied behavior and simplicity makes it a valuable test
environment for fermion simulation techniques \cite{Babbush:2018a}.

Prior work has examined the Fermi-Hubbard model for simulation on quantum computers in
both the Trotter-Suzuki \cite{Dallaire-Demers:2016} and qubitization
\cite{Babbush:2018a} paradigms.  The Trotterized implementation considered a NISQ
architecture, with the controlled-$i$\textsc{swap} as a fundamental gate
\cite{Dallaire-Demers:2016}; their resource estimates, accordingly, will not be
directly comparable with ours. Thus, in Sec.~\ref{sec:trotter-hubbard} we will
make resource estimates both with and without logical Majorana fermion
techniques. The qubitized implementation \cite{Babbush:2018a} considers a similar
fault-tolerant paradigm to the one we work in, however, so we will reference
their resource estimation for comparison to our own.

The qubitized approach to simulation relies on repeated application of a pair of
block-encoding oracles, \textsc{prepare} and \textsc{select}, as specified in
Eq.~\eqref{eq:general-prepare-oracle} and Eq.~\eqref{eq:general-select-oracle}.
These two oracles can be implemented with costs $\bigO(\log N)$ and $10N
+ \bigO(\log N)$ respectively for the $N$-site Fermi-Hubbard model
\cite{Babbush:2018a}, which we will take as an ``industry standard'' to compare
our Majorana-fermion-inspired block encoding oracles to in
Sec.~\ref{sec:qubitized-hubbard}.

%% file: Results.tex
%%%%%%%%%%%%%%%%%%%%%%%%%%%%%%%%%%%%%%%%%%%%%%%%%%%%%%%%%%%%%%%%%%%%%%%%%%%%%%%%
% File: Results.tex
%
% Authors: Andrew J. Landahl <alandahl@sandia.gov>
%          Benjamin C. A. Morrison <benmorr@sandia.gov>
%
%%%%%%%%%%%%%%%%%%%%%%%%%%%%%%%%%%%%%%%%%%%%%%%%%%%%%%%%%%%%%%%%%%%%%%%%%%%%%%%%

%%%%%%%%%%%%%%%%%%%%%%%%%%%%%%%%%%%%%%%%%%%%%%%%%%%%%%%%%%%%%%%%%%%%%%%%%%%%%%%
% Section
%
\section{Results}
\label{sec:logical-Majorana-operations}

Surface code patches can encode logical qubits via the inclusion of twist
defects in their structure.  However, when acting on those twist defects
directly rather than via the abstraction of logical qubit constructions, they
support Majorana fermionic operations; that is, the surface code patches encode
logical Majorana fermions in their twist defects, and it is natural to think of
them as logical Majorana codes, as described in
Sec.~\ref{sec:qubit-surface-code-patches}.  Here, in Sec.~\ref{sec:UQC-LMF}, we
show how one can realize a universal set of operations drawn from the fermionic
quantum circuit gate basis described in
Sec.~\ref{sec:Majorana-fermionic-quantum-circuits} on these logical Majorana
fermions.  Logical Majorana fermions can then serve as a new logical data type,
together with its own set of logical operations, for use in fault-tolerant
fermionic simulation algorithms.

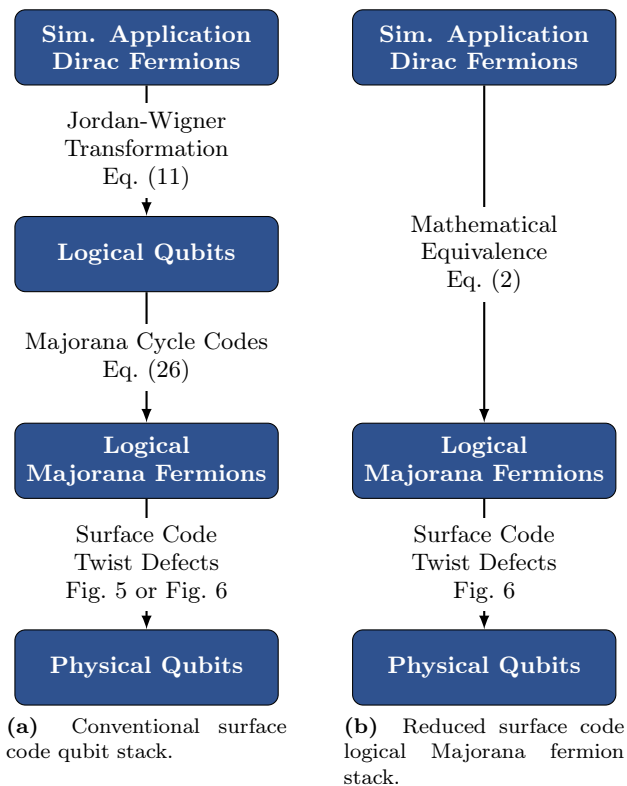
\begin{figure}[H]
\definecolor{sblue}{rgb}{.184,.322,.561}
\tikzstyle{box} = [rectangle, rounded corners, minimum width=3.5cm, minimum height=1cm, align=center, draw=black, fill=sblue, text=white]
\tikzstyle{arrow} = [thick,->,>=latex]
\tikzstyle{arrowlabel} = [align=center,fill=white]
\center{
  \subfigure[\label{fig:old-stack}\
Conventional surface code qubit stack.]{
\begin{tikzpicture}[node distance=2.75cm and 4cm, on grid]
\node (simulation_left) [box] {\textbf{Sim. Application}\\\textbf{Dirac Fermions}};
\node (logical_qubits) [box, below=of simulation_left] {\textbf{Logical Qubits}};
\node (fermions_left) [box, below=of logical_qubits] {\textbf{Logical}\\\textbf{Majorana Fermions}};
\node (physical_left) [box, below=of fermions_left] {\textbf{Physical Qubits}};

\draw [arrow] (simulation_left) -- node[arrowlabel] {Jordan-Wigner\\Transformation\\Eq.~\eqref{eq:fermion-jordan-wigner-transformation}} (logical_qubits);
\draw [arrow] (logical_qubits) -- node[arrowlabel] {Majorana Cycle Codes\\Eq.~\eqref{eq:xbar-zbar-ennon}} (fermions_left);
\draw [arrow] (fermions_left) -- node[arrowlabel] {Surface Code\\Twist Defects\\Fig.~\ref{fig:tetron-hexon-surface-codes} or Fig.~\ref{fig:xxyzz-defect}} (physical_left);
\end{tikzpicture}
}
\qquad
  \subfigure[\label{fig:new-stack}\
Reduced surface code logical Majorana fermion stack.]{
\begin{tikzpicture}[node distance=2.75cm and 4cm, on grid]
\node (simulation_right) [box] {\textbf{Sim. Application}\\\textbf{Dirac Fermions}};
\node (spacer) [shape=rectangle, below=of simulation_right] {};
\node (fermions_right) [box, below=of spacer] {\textbf{Logical}\\\textbf{Majorana Fermions}};
\node (physical_right) [box, below=of fermions_right] {\textbf{Physical Qubits}};

\draw [arrow] (simulation_right) -- node[arrowlabel] {Mathematical\\Equivalence\\Eq.~\eqref{eq:Dirac-to-Majorana}} (fermions_right);
\draw [arrow] (fermions_right) -- node[arrowlabel] {Surface Code\\Twist Defects\\Fig.~\ref{fig:xxyzz-defect}} (physical_right);
\end{tikzpicture}
}
}
\caption{\label{fig:architecture-stack}The removal of an abstraction layer from the architecture stack used in error-corrected simulation applications.}
\end{figure}

Many applications of quantum computing require the implementation of
Dirac fermionic operators. Traditionally, the Jordan-Wigner or another encoding
is used to encode Dirac fermions into qubits (Sec.~\ref{sec:jordan-wigner}). On
a surface-code-based fault-tolerant quantum computer, this means that there is
a stack of abstraction layers from Dirac fermions to logical qubits to surface
code twist defects to physical data qubits, shown in Fig.~\ref{fig:old-stack}.
The twist defect layer is usually elided from descriptions of the surface code
architecture. However, as described in
Sec.~\ref{sec:qubit-surface-code-patches}, the color-changes on the boundaries
of surface code patches have similar properties to internal twist defects and so
are called \emph{corner twists} \cite{Bombin:2010b, Kitaev:2012a, You:2012a,
You:2013a, Yoder:2017a, Brown:2017a, Kesselring:2018a, Brown:2020a,
Lavasani:2018a, Barkeshli:2019a, Zhu:2020a, Bombin:2021a}, and in turn act like
logical Majorana fermions \cite{Bombin:2010b, Brown:2017a, Kesselring:2018a}.
The surface code is in effect a concatenated code with logical qubits encoded in
corner twist Majorana fermions.

As noted in Sec.~\ref{sec:Majorana fermions}, Dirac fermions can be mapped onto
pairs of Majorana fermions. Because twist defects act as logical Majorana
fermions, one can remove one layer of abstraction, replacing both the Majorana
cycle code (Sec.~\ref{sec:Majorana-cycle-codes}) and the Jordan-Wigner
transformation (Sec.~\ref{sec:jordan-wigner}) with the mapping in
Eq.~\eqref{eq:Dirac-to-Majorana}. This creates the simplified simulation
architecture stack in Fig.~\ref{fig:new-stack}.

\begin{figure}[h]
\begin{center}
  \includegraphics[width=0.95\columnwidth]{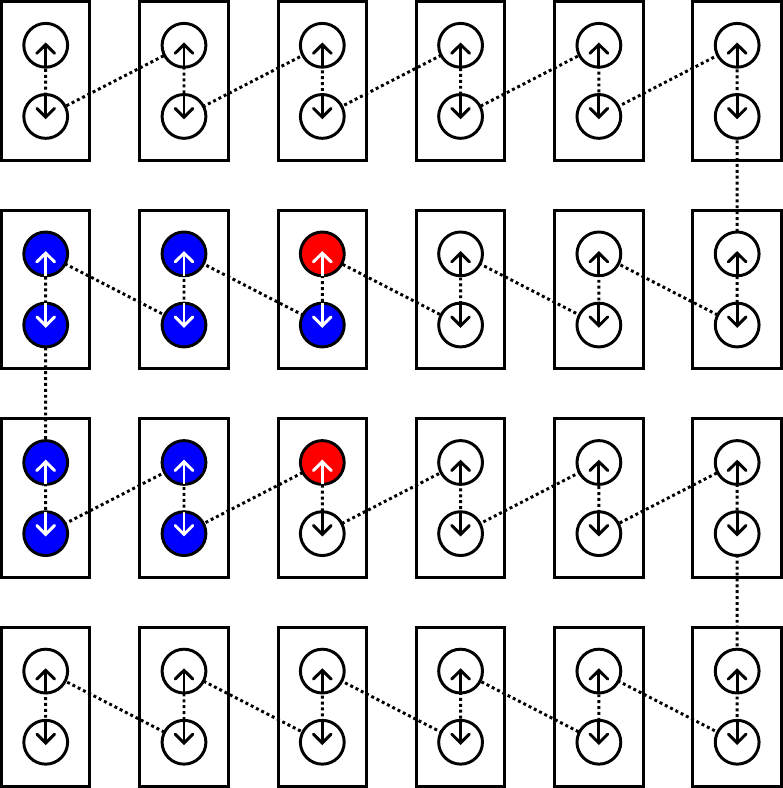}
\caption{\small{\label{fig:JW-snake-vs-2D-lattice} A Jordan-Wigner ``snaking
string'' for the Fermi-Hubbard model.  It travels through two (Dirac) fermions
per site, one spin-up and the other spin-down.  An example nearest-neighbor
hopping interaction
$a_{p,\uparrow}^{\dagger}a\ssc{q,\uparrow}$ between two sites
gives rise to the highly non-local Pauli interaction
$X_{p,\uparrow}\vec{Z}X_{q,\uparrow}
+ Y_{p,\uparrow}\vec{Z}Y_{q,\uparrow}$ depicted by the colored set of
(Dirac) Fermi modes.
}}
\end{center}
\end{figure}

Using this architecture, one can obtain scheduling and parallelization
improvements to fermionic simulation algorithms by mapping local fermionic
operations onto local logical fermionic operations, rather than onto highly
nonlocal Jordan-Wigner strings on logical qubits, as depicted in
Fig.~\ref{fig:JW-snake-vs-2D-lattice}.

Obtaining these improvements typically also involves some understanding of the
system being simulated.  For example, in Sec.~\ref{sec:trotter-hubbard}, to
fully exploit the square-lattice geometry of the fermionic 2D Fermi-Hubbard model, we
move the corner twists of multiple surface-code patches, encoded as described in
Sec.~\ref{sec:qubit-surface-code-patches}, into interior twist defects arrayed in
a 2D square lattice inside a single surface-code patch, as described in
Sec.~\ref{sec:ft-sc}.  This allows us to reduce the asymptotic
Trotter-Suzuki quantum circuit depth from $\bigO(\sqrt{N})$ in a Jordan-Wigner
encoding as depicted in Fig.~\ref{fig:JW-snake-vs-2D-lattice} to a depth of
$\bigO(1)$.  We describe this improvement in more detail in the upcoming
Sec.~\ref{sec:trotter-hubbard}, and we defer a description of how to
interconvert between the different types of relevant surface codes to
Appendix~\ref{sec:code-deformation}.

Sometimes scheduling and parallelization improvements simply become more
manifest in a direct mapping to logical fermions.  For example, in
Sec.~\ref{sec:qubitized-hubbard}, we describe how the \textsc{select} operation in
a qubitized approach to fermionic simulation (described in
Sec.~\ref{sec:qubitization-background}), when applied to the 2D
Fermi-Hubbard model, can be naturally optimized to reduce the $T$-count complexity of
the algorithm by 20\%, from $10N$ to $8N$, where $N$ is the number of sites in
the model.  This optimization can even be applied (but less obviously so) in
a more traditional Jordan-Wigner encoding.  For this reason, we call this
a \emph{Majorana-inspired} speedup.

%%%%%%%%%%%%%%%%%%%%%%%%%%%%%%%%%%%%%%%%%%%%%%%%%%%%%%%%%%%%%%%%%%%%%
% Subsection
%
\subsection{Logical measurement-based topological quantum computation}
\label{sec:UQC-LMF}

As noted in Sec.~\ref{sec:Majorana-fermionic-quantum-circuits}, it is possible
to realize universal quantum computation on Majorana fermionic quantum circuits
solely through preparation and measurement operations on pairs and quads of
Majorana fermions.  Because of the relationship between Majorana fermions and
Ising anyons, which are topological particles, this is essentially
a representation of universal measurement-based topological quantum computation
(MBTQC).  In this section, we demonstrate how to realize MBTQC with the logical
Majorana fermions represented by twist defects in surface codes, when the
surface codes are interpreted as logical Majorana codes instead of logical qubit
codes.  While the majority of the fault-tolerant operations we describe follow
from known constructions on logical qubits stored in surface codes, we are not
aware of them being combined together to form a universal set of gates directly
on the logical Majorana fermions.

In principle, one can construct a parsimonious version of universal logical
MBTQC on surface codes solely using corner twists and lattice surgery operations
on tetron and hexon patches.  However, as we will see in
Secs.~\ref{sec:trotter-hubbard} and~\ref{sec:qubitized-hubbard} (and
elaborated more in Appendix~\ref{sec:code-deformation}), in the context of
speedups to fault-tolerant fermionic quantum simulation algorithms, it is
valuable to allow the corner twists to be brought to the interior so that
fermionic quantum simulations can be accelerated.  These interior twists can
then be braided around one another by sequences of internal surface-code
deformations.

To realize logical MBTQC, as a first step, one must map even-weight logical
Majorana operators onto the physical qubit operators that encode them.  These
operators are the targets of logical state preparations and measurements.
Again, although a parsimonious construction only requires logical Majorana
operators of weight two and four (quadratic and quartic Majorana operators) as
described in Sec.~\ref{sec:Majorana-fermionic-quantum-circuits}, we will show
how to represent any even-weight operator, which opens up the possibility of
further computational speedups in the context of fault-tolerant fermionic
quantum simulation.

\begin{figure}[H]
\center{
  \subfigure[\ Quadratic logical Majorana
  operator.]{\includegraphics[height=1.0in]{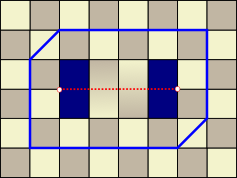}}
\qquad
  \subfigure[\ Global logical Majorana operator (fermionic parity).]{\includegraphics[height=1.0in]{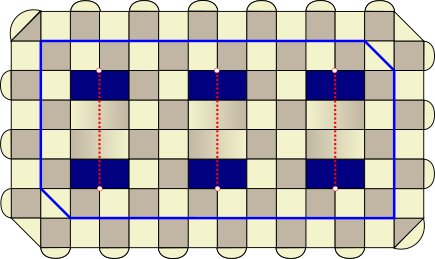}}
}
\caption{\label{fig:majorana-pauli-strings}(Color online.) Logical Majorana
fermion operators encoded as physical qubit Pauli strings.}
\end{figure}

The representation of logical Majorana operators is straightforward: any string of
physical Pauli operators on the qubits that encircles an even number of twists
encodes the even-weight logical Majorana operator supported on those twists.
Furthermore, the product of all of the logical Majorana operators on
a surface-code patch is in the stabilizer group of the code, because it is just
a representation of the conservation of total (logical) fermionic parity.  If
all corner twists are brought to the interior, a Pauli string that encircles
them all can be pulled to the boundary and made to disappear, because there is
nowhere for the string to ``condense'' on the boundary.  This indicates that any
surface-code patch holding only two Majorana fermions stores no information, and
any surface-code patch containing only four Majorana fermions forces those
fermions into a tetron code, or one that can be deformed into one by moving the
twists to the corners; that is, tetron codes encode exactly a single logical
qubit.  Examples of quadratic and global Majorana operators are depicted in
Fig.~\ref{fig:majorana-pauli-strings}.

One of the key differences between universal quantum computation on logical
qubits and logical Majorana fermions represented by surface-code patches is that
there exist even-weight logical Majorana operators that span
multiple patches which have no representation in terms of logical-qubit
operators.  For example, a quadratic Majorana operator in which one twist is on
one patch and one twist is on another has no meaning in terms of logical-qubit
operators.  That said, representing that operator even for logical
Majorana fermions is problematic---two isolated surface code patches have no
reason to be in a state of total fermionic parity conservation between them.  To
represent this operator, one must first deform the two patches so that they
become a single patch.  Only then can the two logical Majorana fermions
``sense'' one another.  We describe how to do this code deformation in detail in
Appendix~\ref{sec:code-deformation}.

Once all twists are on the same patch, the quadratic logical Majorana operators
and the individual logical qubit operators fall into one-to-one correspondence.
In fact, more generally, all even-weight logical Majorana operators correspond
to logical qubit operators that act on one or more logical qubits.  Hence,
fault-tolerant operations for realizing measurements and preparations of these
operators, as described in Sec.~\ref{sec:ft-sc}, can be used to realize these
operations fault-tolerantly.  Although the operations described there acted on
just one or two logical qubits at a time, they generalize straightforwardly to
multi-qubit operations.  For example, one can braid a puncture around
a collection of pairs of interior twists to measure their joint Majorana operator
fault-tolerantly.  Or one can perform multi-logical-qubit lattice surgery with
auxiliary patches to measure an even collection of corner twists.  In essence,
fault-tolerant MBTQC is automatically inherited from fault-tolerant
logical-qubit computation on surface codes, once all the twists are moved to
a single patch.

%%%%%%%%%%%%%%%%%%%%%%%%%%%%%%%%%%%%%%%%%%%%%%%%%%%%%%%%%%%%%%%%%%%%%
% Subsection
%
\subsection{Trotterized Fermi-Hubbard model simulation exemplar}
\label{sec:trotter-hubbard}

Consider a Trotter-Suzuki decomposition of the 2D Fermi-Hubbard model
Hamiltonian on $N$ sites, as represented in
Eq.~\eqref{eq:hubbard-hamiltonian-dirac}.
As noted in Sec.~\ref{sec:trotter-suzuki-background}, to perform this
decomposition, one expresses the Hamiltonian as a sum of Trotter-layer
Hamiltonians, where each such Trotter-layer Hamiltonian is a sum of commuting
terms that do not necessarily commute with the terms in other layers.

In the Jordan-Wigner picture, the qubit Hamiltonian that we seek to Trotterize,
following Babbush \etal\ \cite{Babbush:2018a}, is the following:
%
%\onecolumngrid
%
\begin{align}
\label{eq:hubbard-hamiltonian-jw}
\nonumber
H_\mathrm{HUB} = &-\frac{t}{2}
                  \sum_{\langle p, q \rangle, \sigma}
                  \left( X_{p,\sigma} \vec{Z} X_{q,\sigma} +
                         Y_{p,\sigma} \vec{Z} Y_{q,\sigma} \right)
\\ 
                &+ \frac{u}{4}
                  \sum_{p} Z_{p,\uparrow} Z_{p,\downarrow}
                - \frac{u}{4}
                  \sum_{p,\sigma} Z_{p,\sigma}
                + \frac{uN}{4} \mathds{1}.
\end{align}
%
%\twocolumngrid
%
In this notation, $X$, $Y$, and $Z$ are Pauli matrices acting on qubits at site
$p$ or $q$ as indicated by the first index, and representing fermions with spin
$\sigma \in \{\uparrow,\downarrow\}$ as indicated by the second index.  The
notation $X_{p,\sigma}\vec{Z} X_{q, \sigma}$ is a shorthand for the nonlocal
qubit operator
\begin{align}
X_{p,\sigma} \vec{Z} X_{q,\sigma}
  &:=
  X_{(p,\sigma)}%   \otimes
  Z_{(p,\sigma)+1}% \otimes
  Z_{(p,\sigma)+2}% \otimes
  \cdots% \otimes
  Z_{(q, \sigma)-1}% \otimes
  X_{(q,\sigma)},
\end{align}

where $P_{(p, \sigma)+1}$ denotes the Pauli matrix $P$ acting on the next
combination of site and spin along the Jordan-Wigner string after
$P_(p, \sigma)$, as shown in Fig.~\ref{fig:JW-snake-vs-2D-lattice}.
 
While the final term in this Hamiltonian is proportional to the identity and
therefore has no physical significance, it is usually kept so that the spectrum
of this Hamiltonian matches that of the original Fermi-Hubbard model Hamiltonian.

This can be written as five Trotter layers. Every $Z_{p,\uparrow} Z_{p,\downarrow}$
and $Z_{p,\sigma}$ term commutes with every other such term, so all $Z$ terms can
be combined as a single layer. As for the $P_{p,\sigma}\vec{Z} P_{q, \sigma}$ hopping
terms, where $P \in \{X, Y\}$, they anticommute with any other hopping term with
$\sigma = \sigma'$ and exactly one matching $p$ or $q$. On a square lattice, each
site has hopping terms in four directions, necessitating four different Trotter
layers of hopping terms.

As discussed in Sec.~\ref{sec:trotter-suzuki-background}, this does not
necessarily allow a Trotter step to be implemented in depth 5, or even constant
depth. As a standard architectural assumption, two commuting gates can only be run
in parallel if they act on disjoint sets of qubits. For the the layer of on-site $Z$
terms, this requires decomposition into two sub-layers; all the $Z_{p,\sigma}$ terms
can be implemented simultaneously, but the $Z_{p,\uparrow} Z_{p,\downarrow}$ cannot be
run in parallel with them. The Trotter-Suzuki simulation of the on-site terms, then, can
still be run in $\bigO(1)$ depth.

Because nearest-neighbor fermions in a 2D lattice can be $\bigO(\sqrt{N})$
qubits away in a Jordan-Wigner encoding, as depicted in
Fig.~\ref{fig:JW-snake-vs-2D-lattice}, the ability to parallelize the
quantum simulation of the hopping terms is sharply limited. Each has support on
$\bigO(\sqrt{N})$ qubits, and that bound is tight for at least one term on each site.
By the disjointness requirement, one thus can exponentiate
at most $\Theta(\sqrt{N})$ terms in parallel. Since there are $\Theta(N)$
hopping terms, each Trotter-Suzuki step has a minimum circuit depth (minimum
number of Pauli product sub-layers) of $\Omega(\sqrt{N})$.

In contrast, in the Majorana fermion picture, because each Hamiltonian term is
truly local, one can execute each Trotter-Suzuki step in $\bigO(1)$ depth.
To see this, one first expands each Dirac fermion operator $a$ with spin $\sigma
\in \{\uparrow, \downarrow\}$ at 2D position $p = (x, y)$ into a pair of
Majorana fermions $c_-$ and $c_+$, each of which is given the same spin value
$\sigma$.  After this, the Fermi-Hubbard model Hamiltonian can be expressed solely as
a sum over individual sites as follows:
%
%\onecolumngrid
%
\begin{align}
\label{eq:hubbard-hamiltonian-majorana}
%
%\begin{split}
%
%\nonumber
%
H_\mathrm{HUB}
  = &-\frac{t}{2}
     \sum_{(x, y, \sigma)}
       H^{(\textrm{int})}_{\textrm{HUB}}(x, y, \sigma)
%
%                    \Big[\phantom{+}&c_{x,y,+,\sigma} c_{x+1,y,-,\sigma}
%
%                    \right.
%
%\\[-2.5ex]
%
%\nonumber
%
%                          + c_{x+1,y,+,\sigma} c_{x,y,-,\sigma}
%
%\\
%\\[-2.5ex]
%
%\nonumber
%
%                          + &c_{x,y,+,\sigma} c_{x,y+1,-,\sigma}
%
%\\
%
%\nonumber
%
%                    + &c_{x,y+1,+,\sigma} c_{x,y,-,\sigma}
%
%                    \Big]
%
%\\
%
%\nonumber
%
    +\frac{u}{4}
       \sum_{(x, y)}
         H^{(\textrm{site})}_{\textrm{HUB}}(x, y)
\end{align}
\begin{align}
\nonumber
H^{(\textrm{int})}_{\textrm{HUB}}(x, y, \sigma)
  = &\phantom{+}ic_{x,y,+,\sigma} c_{x+1,y,-,\sigma}
%
%\\
%%
%\nonumber
%
             + ic_{x+1,y,+,\sigma} c_{x,y,-,\sigma}
\\
%
%\nonumber
%
             &+ ic_{x,y,+,\sigma} c_{x,y+1,-,\sigma}
%
%\\
%%
             + ic_{x,y+1,+,\sigma} c_{x,y,-,\sigma}
\\
\nonumber
H^{(\textrm{site})}_{\textrm{HUB}}(x, y)
  = &\phantom{+}i c_{x,y,+,\uparrow} c_{x,y,-,\uparrow}
%
%\\
%%
%\nonumber
%
              + i c_{x,y,+,\downarrow} c_{x,y,-,\downarrow}
\\
             &- c_{x,y,+,\uparrow} c_{x,y,-,\uparrow}
                c_{x,y,+,\downarrow} c_{x,y,-,\downarrow}
              + \mathds{1}
%
%                    \Big[\phantom{+}&i c_{x,y,+,\uparrow} c_{x,y,-,\uparrow}
%
%                    \right.
%
%\\[-2.5ex]
%%
%\nonumber
%%
%                          + &i c_{x,y,+,\downarrow} c_{x,y,-,\downarrow}
%%
%\\
%%
%\nonumber
%%
%                          - &c_{x,y,+,\uparrow} c_{x,y,-,\uparrow} c_{x,y,+,\downarrow} c_{x,y,-,\downarrow}
%%
%\\
%%
%                          + &\mathds{1}
%%
%                    \Big]
%
%\end{split}
\end{align}

Each hopping term is local, whether in the horizontal $x$ direction or vertical
$y$ direction, and is supported on exactly two Majorana fermions; there
are $4N$ such terms, which can be implemented in four Majorana layers, plus
another two layers for the on-site terms, which is very close (asymptotically
identical) to the lower bound of five Trotter layers (the on-site terms commute
with each other but anticommute with every hopping operator, and each Majorana
fermion has four anticommuting hopping operators associated with it).

\begin{figure}[h]
\center{
  \subfigure[\
$c_{x,y,+,\sigma} c_{x,y,-,\sigma}$ (top) and $c_{x,y,+,\uparrow} c_{x,y,-,\uparrow} c_{x,y,+,\downarrow} c_{x,y,-,\downarrow}$ (bottom).]{\includegraphics[width=.2\textwidth]{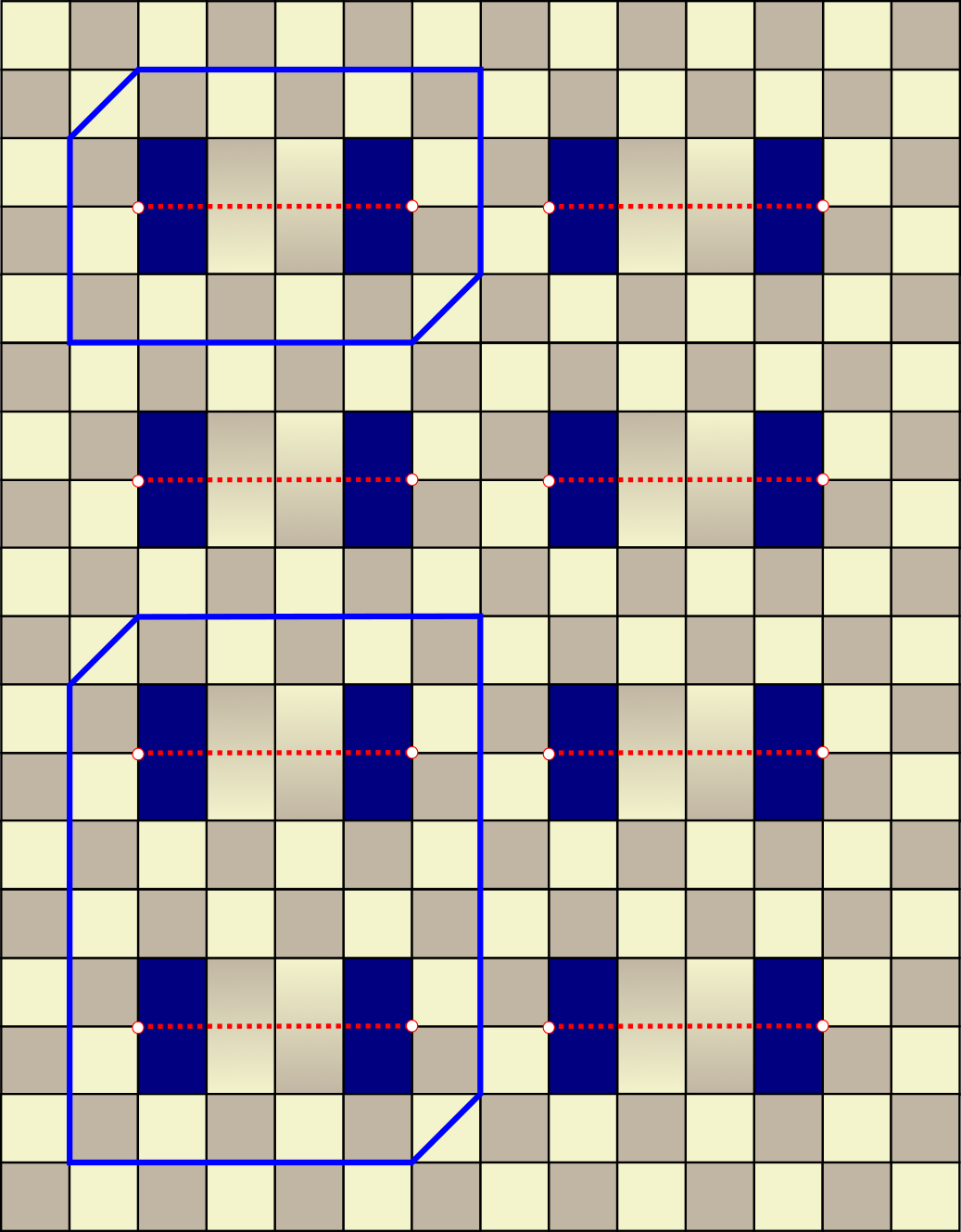}}
\qquad
  \subfigure[\
$c_{x,y,+,\sigma} c_{x+1,y,-,\sigma}$ (top) and $c_{x+1,y,+,\sigma} c_{x,y,-,\sigma}$ (bottom).]{\includegraphics[width=.2\textwidth]{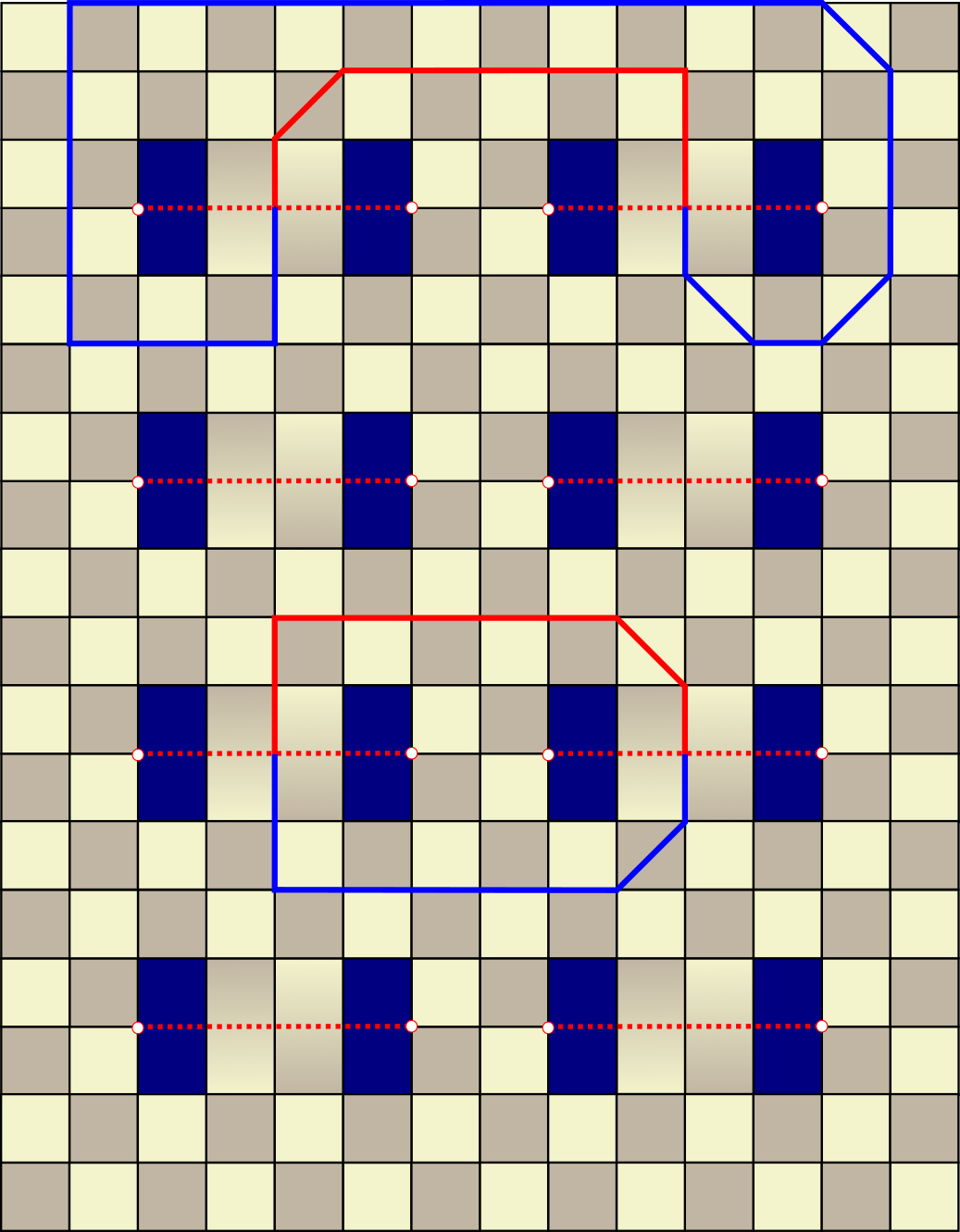}}
\medskip
  \subfigure[\
$c_{x,y,+,\sigma} c_{x,y+1,-,\sigma}$ (left) and $c_{x,y+1,+,\sigma} c_{x,y,-,\sigma}$ (right).]{\includegraphics[width=.2\textwidth]{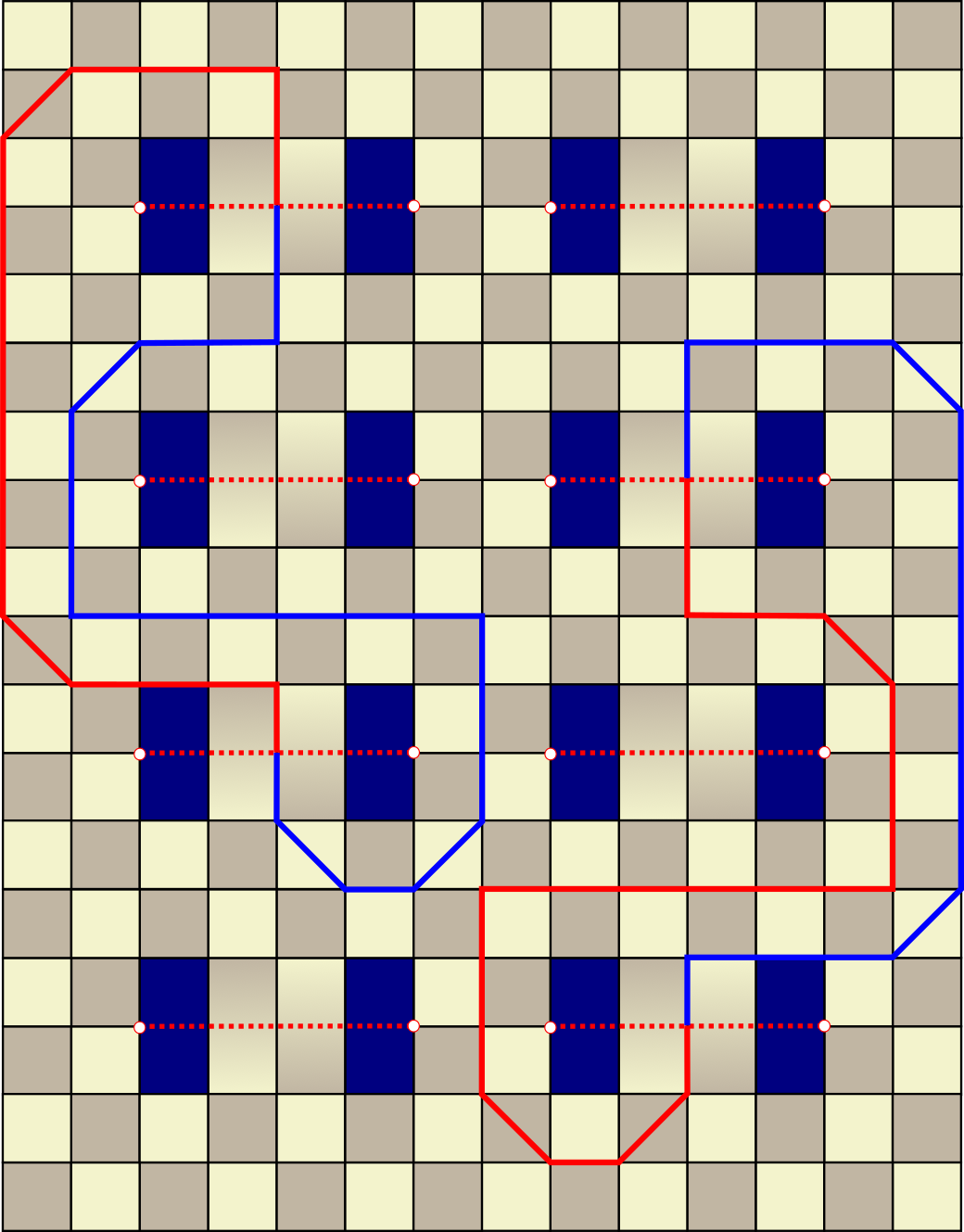}}
\qquad
  \begin{minipage}[t]{.4\textwidth}
%\vspace{-70pt}
  \caption{\label{fig:trotter-twist-grid}The Hamiltonian terms of the
Fermi-Hubbard model, drawn in solid lines as physical Pauli strings around logical
Majorana twists, with $X$ on sites in red and $Z$ on sites in blue. Dark
blue rectangles indicate weight-5 checks, and dashed lines mark missing qubits between
pairs of twists.}
  \end{minipage}
}
\end{figure}

The stringlike logical operators corresponding to each of these terms in
a logical Majorana surface code, in which each of the logical Majoranas is
represented by an internal twist defect in the lattice, is depicted in
Fig.~\ref{fig:trotter-twist-grid}.  One challenge in realizing these stringlike
logical operators is that they sometimes run very close to one another; in order
to maintain a sufficient distance for fault-tolerance with the most compact
packing of twists, first the odd-indexed terms for each layer can be evolved in
parallel, then the even-indexed ones.  Regardless, only a constant gate depth is
required.

%\twocolumngrid

Unlike the speedups described in Sec.~\ref{sec:qubitized-hubbard}, this circuit depth
reduction is actually reliant on access to Majorana fermions encoded as twist defects.
This does not mean it is impossible to describe what is happening in the qubit picture:
applying the Majorana cycle codes in Sec.~\ref{sec:Majorana-cycle-codes} allows the
Majorana operators used to be mapped onto (in general, high-weight) Pauli operators with
the same structure as Jordan-Wigner strings. However, those operators do not have the
same disjoint support as the Majorana operators; and so the parallelization is only
possible if the actual implementation of the exponentiated Hamiltonian terms happens at
the Majorana level.

Furthermore, this construction utilized an encoding in which all twist defects
are on a single patch; in Appendix~\ref{sec:code-deformation}, we show that an
implementation using a collection of tetron-encoded patches that interact via
lattice surgery fails to achieve the same speedup.  This suggests that, for at
least some types of computational advantages, it is important to ensure that
logical fermions are encoded in a way that reflects the layout of the fermions
they are simulating.

%%%%%%%%%%%%%%%%%%%%%%%%%%%%%%%%%%%%%%%%%%%%%%%%%%%%%%%%%%%%%%%%%%%%%
% Subsection
%
\subsection{Majorana-inspired qubitized Fermi-Hubbard model simulation exemplar}
\label{sec:qubitized-hubbard}

Simulating Hamiltonians with the qubitization algorithm requires the
construction of a pair of oracles, \textsc{prepare} and \textsc{select}, as described in
Sec.~\ref{sec:qubitization-background}. For Hamiltonians with many repeated
coefficients, the costs of \textsc{select} dominate.  This is true, for example, for the
Fermi-Hubbard model as studied, \eg, in Ref.~\cite{Babbush:2018a}, for which
\textsc{prepare}
was implemented with $T$ counts logarithmic in the system size $N$, whereas
\textsc{select} had a $T$ count of $10N$ to leading-order.

By choosing our scheme for mapping Hamiltonian terms onto index bitstrings in a way that
accounts for the limited connectivity of our Hamiltonians, more efficient implementations
of \textsc{select} become possible. For example, in the Fermi-Hubbard model, each Hamiltonian term
is supported on, for each spin $\sigma$, either exactly one annihilation and one creation
operator on spin-orbitals with spin $\sigma$; or no fermionic operators with spin $\sigma$
at all. As a result, we can uniquely specify a Hamiltonian term using four indices, each
ranging over a quarter of the possible fermionic operators.

To see a reduction in $T$-count from these indexing scheme modifications, we will also need
a variation on the unary iteration circuits that have been previously used in implementing
\textsc{select} oracles. We refer to this variant of unary iteration, which introduces a parameter
$k$ to iterate over only $1/k$ of the qubits in a register while still applying Pauli
strings across every qubit in the register, as ``stride-$k$ unary iteration,'' and
describe its implementation in Appendix~\ref{sec:stride-unary-iteration}.

These circuits allow us to realize the benefits of the alternate indexing schemes; for
example, in the Fermi-Hubbard model, we can reduce the $T$-count of each
\textsc{select} oracle call
from $10N$ to $8N$ with the new indexing scheme and a stride-2 unary iteration circuit. We
anticipate similar speedups can be realized by reindexing other Hamiltonians with similar
structure.

We want to implement an block-encoding decomposition of the Fermi-Hubbard model Hamiltonian,
Eq.~\eqref{eq:hubbard-hamiltonian-dirac}.
While Babbush \etal\ write this in terms of qubit operators via the
Jordan-Wigner transformation as in Eq.~\eqref{eq:hubbard-hamiltonian-jw},
we will first choose to write it in terms of Majorana operators as
%
%\onecolumngrid
%
\begin{align}
%\begin{split}
%
\nonumber
\sum_\sigma H_{\textrm{HUB}}^{(\textrm{int})}(x, y, \sigma)
  =\phantom{+} &ic_{p,+,\uparrow} \mathds{1}_{+,\downarrow} c_{q,-,\uparrow} \mathds{1}_{-,\downarrow}
\\[-2.5ex]
\nonumber
   + &i\mathds{1}_{+,\uparrow} c_{p,+,\downarrow} \mathds{1}_{-,\uparrow} c_{q,-,\downarrow}
\\
\nonumber
   + &ic_{q,+,\uparrow} \mathds{1}_{+,\downarrow} c_{p,-,\uparrow} \mathds{1}_{-,\downarrow}
\\
   + &i\mathds{1}_{+,\uparrow} c_{q,+,\downarrow} \mathds{1}_{-,\uparrow} c_{p,-,\downarrow}
\end{align}
\begin{align}
\nonumber
\sum_\sigma H_{\textrm{HUB}}^{(\textrm{site})}(x, y)
  = \phantom{+}&i c_{p,+,\uparrow} \mathds{1}_{+,\downarrow} c_{p,-,\uparrow} \mathds{1}_{-,\downarrow}
\\[-2.5ex]
\nonumber
   + &i \mathds{1}_{+,\uparrow} c_{p,+,\downarrow} \mathds{1}_{-,\uparrow} c_{p,-,\downarrow}
\\
   - &c_{p,+,\uparrow} c_{p,+,\downarrow} c_{p,-,\uparrow} c_{p,-,\downarrow}
     + \mathds{1},
%
%\end{split}
\end{align}
%
%\twocolumngrid
%
which differs from Eq.~\eqref{eq:hubbard-hamiltonian-majorana} in Sec.~\ref{sec:trotter-hubbard} only in that we have
explicitly included $\mathds{1}_{\pm,\sigma}$ to draw attention to the fact that each term
has the same form. In the Jordan-Wigner picture, these terms took the forms \(ZZ\), \(Z\),
\(X \vec{Z} X\), and \(Y \vec{Z} Y\); while these do not initially appear to share a
similar structure, the Majorana notation reveals that all four take the form
$c_{j,+,\uparrow} c_{k,+,\downarrow} c_{l,-,\uparrow} c_{m,-,\downarrow}$. In other words,
no term is supported on two of the same type (spin up or down, $+$ or $-$ Majorana) of
Majorana operator on different sites; each is a product of up to one of each of the four.

In order to obtain our speedup, we are going to choose to label our terms with
indices
that correspond to this structure; in particular, we are going to label them with four
$(x, y)$ site coordinates $j$, $k$, $l$, and $m$, each corresponding to one of the four
Majorana operators on that site.

In fact, while the Majorana paradigm made the possibility for this index scheme
clear, we can apply the usual Jordan-Wigner encoding after re-indexing; the
speedup itself is in no way dependent on the logical Majorana fermion
implementation, so we call it a ``Majorana-inspired'' speedup. We write each
Jordan-Wigner string a product of up to one of each of the four types of
fermionic operators $\vec{Z} X_{p,\uparrow}$, $\vec{Z} X_{p,\downarrow}$,
$\vec{Z} Y_{p,\uparrow}$, and $\vec{Z} Y_{p,\downarrow}$, as follows (for \(p
< q\)):
\begin{align} X_{p,\uparrow} \vec{Z} X_{q,\uparrow} &= \left(\vec{Z}
X_{q,\uparrow}\right) \left(\vec{Z} Y_{p,\uparrow}\right) \\ X_{p,\downarrow}
\vec{Z} X_{q,\downarrow} &= \left(\vec{Z} X_{q,\downarrow}\right) \left(\vec{Z}
Y_{p,\downarrow}\right) \\ Y_{p,\uparrow} \vec{Z} Y_{q,\uparrow} &=
\left(\vec{Z} X_{p,\uparrow}\right) \left(\vec{Z} Y_{q,\uparrow}\right) \\
Y_{p,\downarrow} \vec{Z} Y_{q,\downarrow} &= \left(\vec{Z}
X_{p,\downarrow}\right) \left(\vec{Z} Y_{q,\downarrow}\right) \\ Z_{p,\uparrow}
&= \left(\vec{Z} X_{p,\uparrow}\right) \left(\vec{Z} Y_{p,\uparrow}\right) \\
Z_{p,\downarrow} &= \left(\vec{Z} X_{p,\downarrow}\right) \left(\vec{Z}
Y_{p,\downarrow}\right) \\ Z_{p,\uparrow} Z_{p,\downarrow} &= \left(\vec{Z}
X_{p,\uparrow}\right) \left(\vec{Z} Y_{p,\uparrow}\right) \left(\vec{Z}
X_{p,\downarrow}\right) \left(\vec{Z} Y_{p,\downarrow}\right)
\end{align}
Factored in this way, the hidden common structure of the terms is clarified just
as it was in the Majorana perspective.

If one of the four types of fermionic operator is not included in a particular
term, we will have a special index-value for that, denoted here by
\(\varnothing\). Because of this, we will not need dedicated input wires to
\textsc{select} (the \(U\) and \(V\) in Babbush \etal) to denote whether we are
considering a quadratic or quartic Majorana operator term; a quadratic term will
have two \(\varnothing\) indices, and a quartic term will have none.

\onecolumngrid
\begin{align}
\begin{split}
\mathrm{\textsc{prepare}_{HUB}}
\ket{0}^{\otimes 4 + 8 \left\lceil \log\sqrt{\frac{N}{2}+1} \right\rceil}
= \sum_{x=0}^{M-1} \sum_{y=0}^{M-1} & \sqrt{\frac{t}{2\lambda}} \left[
\begin{aligned}
\ket{j_x j_y k_{x+1} k_y l_\varnothing m_\varnothing} & + \ket{j_{x+1} j_y k_x k_y l_\varnothing m_\varnothing} \\
+ \ket{j_x j_y k_x k_{y+1} l_\varnothing m_\varnothing} & + \ket{j_x j_{y+1} k_x k_y l_\varnothing m_\varnothing} \\
+ \ket{j_\varnothing k_\varnothing l_x l_y m_{x+1} m_y} & + \ket{j_\varnothing k_\varnothing l_{x+1} l_y m_x m_y} \\
+ \ket{j_\varnothing k_\varnothing l_x l_y m_x m_{y+1}} & + \ket{j_\varnothing k_\varnothing l_x l_{y+1} m_x m_y}
\end{aligned}
\right] \\
+ \sum_{x=0}^{M-1} \sum_{y=0}^{M-1} & \sqrt{\frac{u}{8\lambda}} \ket{j_x j_y k_x k_y l_x l_y m_x m_y} \\
+ \sum_{x=0}^{M-1} \sum_{y=0}^{M-1} & \sqrt{\frac{u}{4\lambda}} \left[\ket{j_x j_y k_x k_y l_\varnothing m_\varnothing} + \ket{j_\varnothing k_\varnothing l_x l_y m_x m_y}\right]
\end{split}
\end{align}
\twocolumngrid

%\begin{align}
%\begin{split}
%%
%\ket{\mathcal{L}} &=
%\mathrm{\textsc{prepare}_{HUB}}
%%
%\ket{0}^{\otimes 4 + 8 \left\lceil \log\sqrt{\frac{N}{2}+1} \right\rceil} \\
%%
% &= \sum_{x=0}^{M-1} \sum_{y=0}^{M-1} \left[
%\begin{aligned}
%\sqrt{\frac{t}{2\lambda}} \ket{j_x j_y k_{x+1} k_y l_\varnothing m_\varnothing} \\
%+ \sqrt{\frac{t}{2\lambda}} \ket{j_{x+1} j_y k_x k_y l_\varnothing m_\varnothing} \\
%+ \sqrt{\frac{t}{2\lambda}} \ket{j_x j_y k_x k_{y+1} l_\varnothing m_\varnothing} \\
%+ \sqrt{\frac{t}{2\lambda}} \ket{j_x j_{y+1} k_x k_y l_\varnothing m_\varnothing} \\
%+ \sqrt{\frac{t}{2\lambda}} \ket{j_\varnothing k_\varnothing l_x l_y m_{x+1} m_y} \\
%+ \sqrt{\frac{t}{2\lambda}} \ket{j_\varnothing k_\varnothing l_{x+1} l_y m_x m_y} \\
%+ \sqrt{\frac{t}{2\lambda}} \ket{j_\varnothing k_\varnothing l_x l_y m_x m_{y+1}} \\
%+ \sqrt{\frac{t}{2\lambda}} \ket{j_\varnothing k_\varnothing l_x l_{y+1} m_x m_y} \\
%+ \sqrt{\frac{u}{8\lambda}} \ket{j_x j_y k_x k_y l_x l_y m_x m_y} \\
%+ \sqrt{\frac{u}{4\lambda}} \ket{j_x j_y k_x k_y l_\varnothing m_\varnothing} \\
%+ \sqrt{\frac{u}{4\lambda}} \ket{j_\varnothing k_\varnothing l_x l_y m_x m_y}
%\end{aligned}
%\right]
%%
%\end{split}
%\end{align}

We use five auxiliary qubits to prepare this state. The first two flag whether we are
constructing an \(X \vec{Z} X\)/\(Y \vec{Z} Y\) term, a \(ZZ\) term, or a \(Z\) term, in
the same manner as Babbush \etal's \(U\) and \(V\), except that we will not need to pass
them to \textsc{select}. The next three determine which of the eight possible
\(X \vec{Z} X\)/\(Y \vec{Z} Y\) terms, or which of the two possible \(Z\) terms, we
construct, by controlling how the various indices are \textsc{swap}ped around between registers.

\onecolumngrid

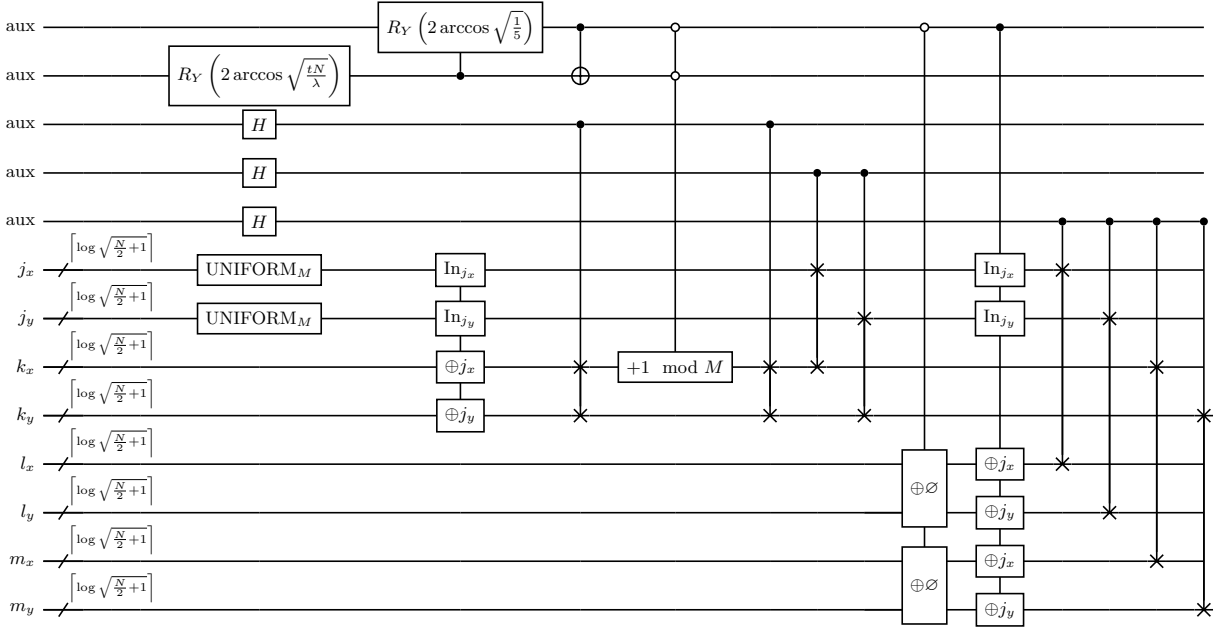
\begin{figure}[H]
\begin{center}
\resizebox{0.9\textwidth}{!}{%
\begin{quantikz}[row sep={0.85cm,between origins}]
\lstick{${\mathrm{aux}}$} &[2mm] \qw & \qw & \qw & \qw & \gate{{R_Y}\left(2
\arccos \sqrt{\frac{1}{5}}\right)} & \ctrl{1} & \octrl{1} & \qw & \qw & \qw & \octrl{9} & \ctrl{5} & \qw & \qw & \qw & \qw \\
\lstick{${\mathrm{aux}}$} &[2mm] \qw & \qw & \qw & \gate{{R_Y}\left(2 \arccos
\sqrt{\frac{tN}{\lambda}}\right)} & \ctrl{-1} & \targ{} & \octrl{6} & \qw & \qw & \qw & \qw & \qw & \qw & \qw & \qw & \qw \\
\lstick{${\mathrm{aux}}$} &[2mm] \qw & \qw & \qw & \gate{{H}} & \qw & \ctrl{6} & \qw & \ctrl{6} & \qw & \qw & \qw & \qw & \qw & \qw & \qw & \qw \\
\lstick{${\mathrm{aux}}$} &[2mm] \qw & \qw & \qw & \gate{{H}} & \qw & \qw & \qw & \qw & \ctrl{4} & \ctrl{5} & \qw & \qw & \qw & \qw & \qw & \qw \\
\lstick{${\mathrm{aux}}$} &[2mm] \qw & \qw & \qw & \gate{{H}} & \qw & \qw & \qw & \qw & \qw & \qw & \qw & \qw & \ctrl{5} & \ctrl{6} & \ctrl{7} & \ctrl{8} \\
\lstick{${j_x}$} &[2mm] \qw\qwbundle{\left\lceil \log \sqrt{\frac{N}{2}+1}
\right\rceil} & \qw & \qw & \gate{\mathrm{UNIFORM}_M} & \gate{\mathrm{In}_{j_x}}\vqw{3} & \qw & \qw & \qw & \swap{2} & \qw & \qw & \gate{\mathrm{In}_{j_x}}\vqw{1} & \swap{4} & \qw & \qw & \qw \\
\lstick{${j_y}$} &[2mm] \qw\qwbundle{\left\lceil \log \sqrt{\frac{N}{2}+1}
\right\rceil} & \qw & \qw & \gate{\mathrm{UNIFORM}_M} & \gate{\mathrm{In}_{j_y}} & \qw & \qw & \qw & \qw & \swap{2} & \qw & \gate{\mathrm{In}_{j_y}}\vqw{3} & \qw & \swap{4} & \qw & \qw \\
\lstick{${k_x}$} &[2mm] \qw\qwbundle{\left\lceil \log \sqrt{\frac{N}{2}+1}
\right\rceil} & \qw & \qw & \qw & \gate{\oplus {j_x}} & \swap{1} & \gate{+1 \mod M} & \swap{1} & \targX {} & \qw & \qw & \qw & \qw & \qw & \swap{4} & \qw \\
\lstick{${k_y}$} &[2mm] \qw\qwbundle{\left\lceil \log \sqrt{\frac{N}{2}+1}
\right\rceil} & \qw & \qw & \qw & \gate{\oplus {j_y}} & \targX{} & \qw & \targX{} & \qw & \targX{} & \qw & \qw & \qw & \qw & \qw & \swap{4} \\
\lstick{${l_x}$} &[2mm] \qw\qwbundle{\left\lceil \log \sqrt{\frac{N}{2}+1}
\right\rceil} & \qw & \qw & \qw & \qw & \qw & \qw & \qw & \qw & \qw & \gate[wires=2]{\oplus \varnothing}\vqw{2} & \gate{\oplus {j_x}}\vqw{1} & \targX{} & \qw & \qw & \qw \\
\lstick{${l_y}$} &[2mm] \qw\qwbundle{\left\lceil \log \sqrt{\frac{N}{2}+1}
\right\rceil} & \qw & \qw & \qw & \qw & \qw & \qw & \qw & \qw & \qw & \qw & \gate{\oplus {j_y}}\vqw{1} & \qw & \targX{} & \qw & \qw \\
\lstick{${m_x}$} &[2mm] \qw\qwbundle{\left\lceil \log \sqrt{\frac{N}{2}+1}
\right\rceil} & \qw & \qw & \qw & \qw & \qw & \qw & \qw & \qw & \qw & \gate[wires=2]{\oplus \varnothing} & \gate{\oplus {j_x}}\vqw{1} & \qw & \qw & \targX{} & \qw \\
\lstick{${m_y}$} &[2mm] \qw\qwbundle{\left\lceil \log \sqrt{\frac{N}{2}+1}
\right\rceil} & \qw & \qw & \qw & \qw & \qw & \qw & \qw & \qw & \qw & \qw & \gate{\oplus {j_y}} & \qw & \qw & \qw & \targX{} \\
\end{quantikz}
}%
\end{center}
\caption{\label{fig:prepare-circuit}The qubitized \textsc{prepare} oracle circuit.}
\end{figure}
\vspace{0\baselineskip}
\twocolumngrid

Once we have prepared the correct index state, our \textsc{select} operator just applies precisely
the Jordan-Wigner-transformed Majorana operators specified by the
index, using the stride-$k$ unary iteration we described in
Appendix~\ref{sec:stride-unary-iteration}. Specifically, we will use the $k=2$ version of the
construction to iterate over only every other lattice site—that is, either only
the spin-up or only the spin-down sites—while still applying a Jordan-Wigner string
operator on every lattice site, and will do so at only the \(T\)-cost of iterating over
every other lattice site (that is, approximately half the \(T\)-cost of iterating over
every lattice site).

\onecolumngrid
%\[
%\begin{aligned}
%\mathrm{SELECT_{HUB}} \ket{jklm}\ket{\psi} = & \ket{jklm} \\
%\otimes &
%\left\{
%    \begin{array}{lr}
%      \mathds{1} & \mathrm{if\ } j = \varnothing \\
%      \vec{Z} X_{j,\uparrow} & 
%    \end{array}
%\right\}
%\left\{
%    \begin{array}{lr}
%      \mathds{1} & \mathrm{if\ } k = \varnothing \\
%      \vec{Z} Y_{k,\uparrow} & 
%    \end{array}
%\right\} \\
%& \left\{
%    \begin{array}{lr}
%      \mathds{1} & \mathrm{if\ } l = \varnothing \\
%      \vec{Z} X_{l,\downarrow} & 
%    \end{array}
%\right\}
%\left\{
%    \begin{array}{lr}
%      \mathds{1} & \mathrm{if\ } m = \varnothing \\
%      \vec{Z} Y_{m,\downarrow} & 
%    \end{array}
%\right\} \ket{\psi}
%\end{aligned}
%\]
%
\begin{align}
\begin{split}
\mathrm{\textsc{select}_{HUB}} \ket{jklm}\ket{\psi} = \ket{jklm} 
\otimes 
\left\{
    \begin{array}{lr}
      \mathds{1} & \mathrm{if\ } j = \varnothing \\
      \vec{Z} X_{j,\uparrow} & 
    \end{array}
\right\}
\left\{
    \begin{array}{lr}
      \mathds{1} & \mathrm{if\ } k = \varnothing \\
      \vec{Z} Y_{k,\uparrow} & 
    \end{array}
\right\} 
\left\{
    \begin{array}{lr}
      \mathds{1} & \mathrm{if\ } l = \varnothing \\
      \vec{Z} X_{l,\downarrow} & 
    \end{array}
\right\}
\left\{
    \begin{array}{lr}
      \mathds{1} & \mathrm{if\ } m = \varnothing \\
      \vec{Z} Y_{m,\downarrow} & 
    \end{array}
\right\} \ket{\psi}
\end{split}
\end{align}

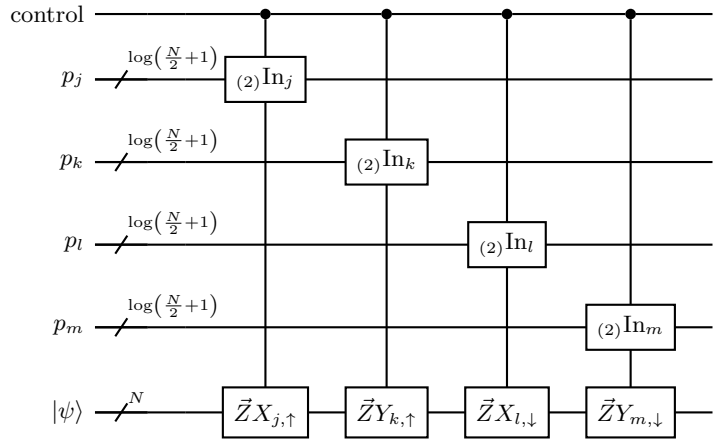
\begin{figure}[H]
\begin{center}
\begin{quantikz}
\lstick{${\mathrm{control}}$} &[2mm] \qw & \qw & \ctrl{1} & \ctrl{2} & \ctrl{3} & \ctrl{4} & \qw \\
\lstick{${p_j}$} &[2mm] \qw\qwbundle{\log \left(\frac{N}{2}+1\right)} & \qw & \gate{_{(2)}\mathrm{In}_j}\vqw{4} & \qw & \qw & \qw & \qw \\
\lstick{${p_k}$} &[2mm] \qw\qwbundle{\log \left(\frac{N}{2}+1\right)} & \qw & \qw & \gate{_{(2)}\mathrm{In}_k}\vqw{3} & \qw & \qw & \qw \\
\lstick{${p_l}$} &[2mm] \qw\qwbundle{\log \left(\frac{N}{2}+1\right)} & \qw & \qw & \qw & \gate{_{(2)}\mathrm{In}_l}\vqw{2} & \qw & \qw \\
\lstick{${p_m}$} &[2mm] \qw\qwbundle{\log \left(\frac{N}{2}+1\right)} & \qw & \qw & \qw & \qw & \gate{_{(2)}\mathrm{In}_m}\vqw{1} & \qw \\
\lstick{$\ket{\psi}$} &[2mm] \qw\qwbundle{N} & \qw & \gate{\vec{Z} X_{j,\uparrow}} & \gate{\vec{Z} Y_{k,\uparrow}} & \gate{\vec{Z} X_{l,\downarrow}} & \gate{\vec{Z} Y_{m,\downarrow}} & \qw
\end{quantikz}
\end{center}
\caption{\label{fig:select-circuit}The qubitized \textsc{select} oracle circuit.}
\end{figure}
\vspace{0\baselineskip}
\twocolumngrid

Each of the four unary iterations above iterates over \(\frac{N}{2}+1\) indices, and thus
has a $T$-count of \(2N\). In total, this means our \textsc{select}
implementation has a $T$-count
of \(8 N + \bigO(\log N) \), compared to Babbush \etal's
\(10 N + \bigO(\log N)\).

%% file: Conclusion.tex
%%%%%%%%%%%%%%%%%%%%%%%%%%%%%%%%%%%%%%%%%%%%%%%%%%%%%%%%%%%%%%%%%%%%%%%%%%%%%%%
% File: Conclusion.tex
% 
% Authors: Andrew J. Landahl <alandahl@sandia.gov>
%          Benjamin C. A. Morrison <benmorr@sandia.gov>
%
%%%%%%%%%%%%%%%%%%%%%%%%%%%%%%%%%%%%%%%%%%%%%%%%%%%%%%%%%%%%%%%%%%%%%%%%%%%%%%%

%%%%%%%%%%%%%%%%%%%%%%%%%%%%%%%%%%%%%%%%%%%%%%%%%%%%%%%%%%%%%%%%%%%%%%%%%%%%%%%
%Section
%
\section{Conclusion}
\label{sec:conclusion}

Inspired by Feynman's call to simulate quantum mechanics with quantum
computers to eliminate the overhead in quantum-to-classical mappings
\cite{Feynman:1982a}, we have shown how to simulate fermions with
error-corrected logical Majorana fermions in a way that eliminates the
overhead in fermion-to-qubit mappings.  We did so by processing the logical
Majorana fermions stored as twist defects in surface code patches.
We rely on known constructions on surface codes to implement the required
operations for this processing, most of which have not been previously
applied to the target of logical Majorana fermionic computation, but to the
processing of logical qubits in Majorana cycle codes.
We expect that these
constructions will apply to topological codes more broadly than just the
surface code, including those defined generally by rotation systems
\cite{Sarkar:2021a} and those defined dynamically, such as the honeycomb
code \cite{Hastings:2021a, Gidney:2021a}.

We demonstrate the value of the ability to manipulate logical Majorana
fermions directly in a fault-tolerant setting for optimizing quantum simulation
algorithms in the exemplar system of a 2D Fermi-Hubbard model. The improvements
in this exemplar suggest that similar optimizations may be possible in other
quantum simulation algorithms. In particular, removing elaborate nonlocal
sequences of CNOT gates used to facilitate operations in the Jordan-Wigner
mapping of qubits to fermions, such as those described in Refs.~\cite{Babbush:2018a, vonBurg:2021a, Lee:2021a, Wan:2021a}, may lead to
opportunities for parallel execution and scheduling optimizations.

Furthermore, we show the value of the logical fermionic data type as a conceptual
tool to aid in quantum software development. By working in the abstraction of
fermionic computation, we are able to obtain a $T$-count reduction for block
encoding oracles in the same exemplar system, which is applicable even in
non-fermionic architectures. We expect this approach to be a useful
aid to algorithm developers examining other target applications as well.

Developing optimizing compilers that exploit the
availability of native (Majorana) fermionic operations is an interesting
avenue for further research.
We look forward to seeing how the broad array of
fermionic quantum simulation applications will be able to exploit the
elimination of fermion-to-qubit mappings that we describe here.

As shown in Ref.~\cite{Kesselring:2018a}, one can realize a rich panoply of
anyons as twist defects in topological codes, even beyond the simple Majorana
fermions we have considered.  This suggests that our approach is extendable in
a way that facilitates low-overhead fault-tolerant anyonic simulation generally,
using these logical anyons in the measurement-based topological quantum
computing paradigm.  For example, the Fradkin-Kadanoff transformation for
parafermions that generalizes the Jordan-Wigner transformation for fermions
might be able to be eliminated to facilitate less resource intensive studies of
parafermions with quantum computers \cite{Liu:2021a, Fradkin:1980a}.
Simulations of anyonic physics might even help to develop technology based on
actual anyonic excitations in material systems \cite{Beenakker:2011a}.

While our approach may facilitate simulation studies of anyonic physics,
including of Majorana fermions themselves, the fact that our constructions
allow one to manipulate arbitrarily reliable ``synthetic'' logical Majorana
fermions directly suggests that our approach could be an alternative to
manipulating Majorana fermions realized as quasiparticle excitations in
condensed matter systems for the purposes of reliable quantum computation.
Much of the effort developed for how to manipulate Majorana fermions for the
purposes of quantum computation, for example protocols for ``topological
quantum compiling,'' \cite{Kliuchnikov:2014a} can be mapped to the logical
Majorana fermion setting without any modifications.  That said, the quest to
realize physical Majorana fermions is still very important for fundamental
physics and could be enabling for some quantum technologies.

It is worth noting that tailoring quantum error correcting codes for
explicit use in fermionic quantum simulation algorithms is not a new idea;
for example, see Refs.~\cite{Jiang:2018a, Steudtner:2019a, Setia:2019a,
Derby:2020a, Jiang:2020a}.  However, all previous constructions of which we
are aware either used \textit{ad hoc} codes or only worked at fixed code
distances which did not facilitate arbitrarily reliable quantum simulations.
By basing our approach on surface codes, which have been studied
extensively, our constructions will work at arbitrary code distances and
could be realized by technologies built around surface codes.

%% file: Acknowledgments.tex
%%%%%%%%%%%%%%%%%%%%%%%%%%%%%%%%%%%%%%%%%%%%%%%%%%%%%%%%%%%%%%%%%%%%%%%%%%%%%%%%
% File: Acknowledgments.tex
%
% Authors: Andrew J. Landahl <alandahl@sandia.gov>
%          Benjamin C. A. Morrison <benmorr@sandia.gov>
%
%%%%%%%%%%%%%%%%%%%%%%%%%%%%%%%%%%%%%%%%%%%%%%%%%%%%%%%%%%%%%%%%%%%%%%%%%%%%%%%%

%%%%%%%%%%%%%%%%%%%%%%%%%%%%%%%%%%%%%%%%%%%%%%%%%%%%%%%%%%%%%%%%%%%%%%%%%%%%
% Acknowledgments

This paper benefited from helpful conversations from many people, including (in
alphabetical order) Andrew Baczewski, Benjamin Brown, Riley Chien, Anand Ganti,
Markus Kesselring, Daniel Litinski, Jesse Lutz, Setso Metodi, Stefan Seritan,
Jaimie Stephens, James Whitfield, and Yijia Xu.  We would like to thank the following
people for their helpful feedback on an early draft of this paper: Andrew
Baczewski, Lucas Kocia, Kenny Rudinger, Stefan Seritan, and Wayne Witzel.
Finally, we thank Daniel Litinski, for inspiring the color scheme we used to
depict surface codes.

This material is based upon work supported by the U.S.\ Department of Energy,
Office of Science, National Quantum Information Science Research Centers.
Additional support is acknowledged from National Nuclear Security
Administration's Advanced Simulation and Computing Program, and the Laboratory
Directed Research and Development program at Sandia National Laboratories.

Sandia National Laboratories is a multimission laboratory managed and
operated by National Technology and Engineering Solutions of Sandia, LLC., a
wholly owned subsidiary of Honeywell International, Inc., for the U.S.\
Department of Energy's National Nuclear Security Administration under
contract DE-NA-0003525.
  
This paper describes objective technical results and analysis. Any
subjective views or opinions that might be expressed in the paper do not
necessarily represent the views of the U.S.\ Department of Energy or the
United States Government.

%% file: Appendix.tex
%%%%%%%%%%%%%%%%%%%%%%%%%%%%%%%%%%%%%%%%%%%%%%%%%%%%%%%%%%%%%%%%%%%%%%%%%%%%%%%%
% File: Appendix.tex
%
% Authors: Andrew J. Landahl <alandahl@sandia.gov>
%          Benjamin C. A. Morrison <benmorr@sandia.gov>
%
%%%%%%%%%%%%%%%%%%%%%%%%%%%%%%%%%%%%%%%%%%%%%%%%%%%%%%%%%%%%%%%%%%%%%%%%%%%%%%%%

%%%%%%%%%%%%%%%%%%%%%%%%%%%%%%%%%%%%%%%%%%%%%%%%%%%%%%%%%%%%%%%%%%%%%%%%%%%%
% Appendix

%%%%%%%%%%%%%%%%%%%%%%%%%%%%%%%%%%%%%%%%%%%%%%%%%%%%%%%%%%%%%%%%%%%%%
% Section
%
\section{Unary iteration circuits}
\label{sec:unary-iteration-background}

The block-encoding oracles for the Hubbard model in Ref.~\cite{Babbush:2018a}, as mentioned in
Sec.~\ref{sec:hubbard-model-background}, rely on the quantum algorithmic primitive of
unary iteration \cite{Gidney:2019}. Our block-encoding oracles in
Sec.~\ref{sec:qubitized-hubbard} will use a modified version of that primitive, which we
construct in Appendix~\ref{sec:stride-unary-iteration}. We briefly describe here the standard
version of unary iteration, for comparison.

In general, the unary iteration primitive implements the operation
\begin{align}
\ket{c}\ket{l}\ket{\psi} \rightarrow \ket{c}\ket{l} \Lambda_c(\hat{A}_l)
\ket{\psi}
\end{align}
for a collection of operations $\hat{A}_0 ... \hat{A}_{L-1}$ with $T$-count
$4L-4$ plus any $T$-cost for implementing the $A_l$, where $\Lambda_c(U)$
denotes the controlled-$U$ operation, controlled on the state of the qubit
$|c\rangle$. The specific implementation
used in the \textsc{select} oracles of Ref.~\cite{Babbush:2018a} specializes
this to ``ranged and indexed'' operations
\begin{align}
\ket{c}\ket{l}\ket{\psi} \rightarrow \ket{c}\ket{l} \Lambda_c\left(
\bigotimes_{i=0}^{l-1} \hat{P}_i \otimes \hat{Q}_l \right) \ket{\psi}
\end{align}
where $P_i$ and $Q_i \in {X_i, Y_i, Z_i, I_i}$ are Pauli operators acting on
qubit $i$ of the $\psi$ register. This is done by the following circuit
construction:

\onecolumngrid

\begin{figure}[H]
\begin{center}
\resizebox{0.9\textwidth}{!}{%
\begin{quantikz}[row sep={0.8cm,between origins}]
\lstick{${\mathrm{control}}$} &[2mm] \ctrl{5} & \ctrl{1} & \qw & \qw & \qw & \qw & \qw & \ctrl{2} & \qw & \qw & \qw & \qw & \qw & \qw & \ctrl{1} & \qw \\
\lstick{${l_1}$} &[2mm] \qw & \octrl{-1} & \qw & \qw & \qw & \qw & \qw & \qw & \qw & \qw & \qw & \qw & \qw & \qw & \ctrl{-1} & \qw \\
&[2mm] & \vqw{-1} & \ctrl{1} & \qw & \ctrl{2} & \qw & \ctrl{1} & \targ{} & \ctrl{1} & \qw & \ctrl{2} & \qw & \qw & \ctrl{1} & \qw\vqw{-1} \\
\lstick{${l_0}$} &[2mm] \qw & \qw & \octrl{-1} & \qw & \qw &\qw & \ctrl{-1} & \qw & \octrl{-1} & \qw & \qw & \qw & \qw & \ctrl{-1} & \qw & \qw \\
&[2mm] & & \vqw{-1} & \ctrl{2} & \targ{} & \ctrl{3} & \qw\vqw{-1} & & \vqw{-1} & \ctrl{4} & \targ{} & \ctrl{1} & \ctrl{5} & \qw\vqw{-1} \\
\lstick{${\mathrm{accumulator}}$} &[2mm] \vqw{-5} & \qw & \qw & \targ{} & \ctrl{1} & \targ{} & \ctrl{2} & \qw & \qw & \targ{} & \ctrl{3} & \qw\vqw{-1} & \\
\lstick{${\ket{\psi}_{00}}$} &[2mm] \qw & \qw & \qw & \gate{Q} & \gate{P} & \qw & \qw & \qw & \qw & \qw & \qw & \qw & \qw & \qw & \qw & \qw \\
\lstick{${\ket{\psi}_{01}}$} &[2mm] \qw & \qw & \qw & \qw & \qw & \gate{Q} & \gate{P} & \qw & \qw & \qw & \qw & \qw & \qw & \qw & \qw & \qw \\
\lstick{${\ket{\psi}_{10}}$} &[2mm] \qw & \qw & \qw & \qw & \qw & \qw & \qw & \qw & \qw & \gate{Q} & \gate{P} & \qw & \qw & \qw & \qw & \qw \\
\lstick{${\ket{\psi}_{11}}$} &[2mm] \qw & \qw & \qw & \qw & \qw & \qw & \qw & \qw & \qw & \qw & \qw & \qw & \gate{Q} & \qw & \qw & \qw \\
\end{quantikz}
}%
\caption{\label{fig:unary-ranged-indexed}A general ranged and indexed operator implemented
using unary iteration.}
\end{center}
\end{figure}
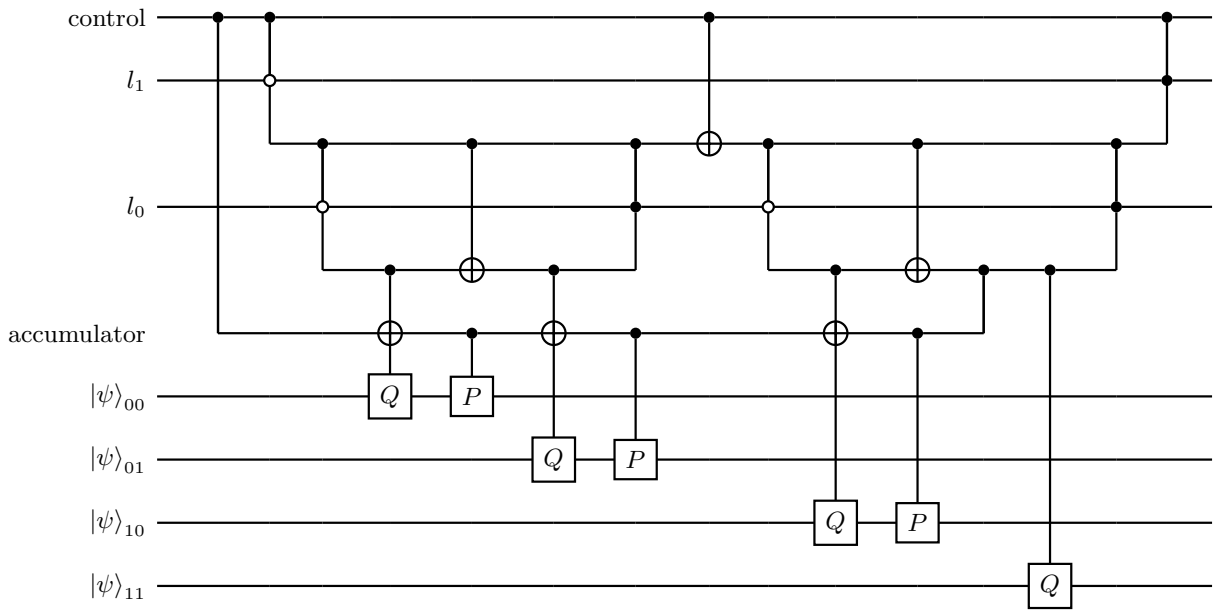

\twocolumngrid

As noted in Ref.~\cite{Babbush:2018a}, this construction can be transformed
much as classical iterators can; in Appendix~\ref{sec:stride-unary-iteration} we discuss a
transformation of particular utility for our work.

%%%%%%%%%%%%%%%%%%%%%%%%%%%%%%%%%%%%%%%%%%%%%%%%%%%%%%%%%%%%%%%%%%%%%
% Section
%
\section{Code deformation between the $n$-tetron and $(2n+2)$-on encodings}
\label{sec:code-deformation}

As we noted in Sec.~\ref{sec:UQC-LMF}, a quadratic logical Majorana fermion
operator that spans two different tetron surface-code patches has no
representation in terms of logical-qubit operators.  In order for this operator
to have meaning, even for logical Majorana fermions, the twists where they
reside need to be brought onto the same surface-code patch so that they can
``sense'' one another through a global fermionic parity constraint (which is
equal to the product of all the stabilizer generators of the collective patch).
This is necessary, generically, for fault-tolerant fermionic quantum simulation
algorithms that exploit logical Majorana fermions directly, because, generally,
a fermionic Hamiltonian can include quadratic operators between any pair of
logical Majorana fermions.

That said, one might wish to store logical Majorana fermions on tetron patches
for ease of implementation; one might imagine that one could transform the
information carriers between the $(2n+2)$-fermion Majorana cycle code and and an
array of $n$ tetron codes on demand. Unfortunately, we will see that, while this
is possible, it is sufficiently nonlocal and expensive as to obviate any
locality-derived scheduling improvements one might have obtained.

Even if we are not intending to use our logical Majorana fermions to implement
logical qubits, the operators described in described in
Sec.~\ref{sec:Majorana-cycle-codes} for logical Pauli operators form a complete
set of generators for the group of logical Majorana fermion operators available
within the code (that is, excluding any that anticommute with stabilizer-group
elements---most notably the odd-weight operators---and treating each operator as
its equivalence class under stabilizer-element multiplication). Specifically,
the $n$-tetron code has the same stabilizer generators as are enforced by $n$
four-logical-Majorana-fermion square patches, and the
$(2n+2)$-logical-Majorana-fermion cycle code has the same stabilizer generators
as the dislocation twist defect code. Accordingly, we can use the stabilizer
formalism \cite{Gottesman:1997a} to track the code deformations we perform.

Suppose we wish to deform from a cycle code with $2n+2$ logical Majorana
fermions on a single patch to a code with patches of four logical Majorana
fermions each. We consider two different versions of this deformation. In the
first, only defined when $n$ is odd, we keep the fermion number constant, and
split the large patch into $\frac{n+1}{2}$ tetron patches. However, adding
$\frac{n-1}{2}$ additional stabilizer generators but keeping the number of
information carriers constant will shrink our logical subspace, and we will lose
information.  Instead, we will need to add $2n-2$ additional logical Majorana
fermions, leaving us with a total of $4n$ logical Majorana fermions on $n$
tetron patches. Thus, we will have added precisely half as many checks as
logical Majorana fermions, and so the dimension of our logical subspace remains
constant.

For ease of description, we use four additional auxiliary logical Majorana
fermions during the deformation; they can be optimized away, but doing so makes
the operation less clear.  Suppose we have a linear array of $2n+2$ Majorana
fermions $c_0 ... c_{2n+1}$ on a single patch. Our logical operator generators
are
\begin{align}
L_k &= \prod_{i=0}^{k} c_{2i} c_{2i+1} \\ L'_k &= c_{2k+1} c_{2k+2}
\end{align}
for $0 \le k < n$ as described above. Our only stabilizer generator is the
product
\begin{align}
S_G = \prod_{i=0}^{2n+1} c_i
\end{align}
of all the logical Majorana operators (that is, global parity conservation).
\begin{figure}[H]
\center{
  \includegraphics[height=0.375in]{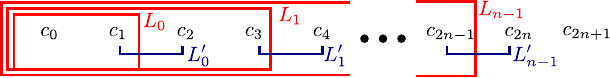}
}
\caption{\label{fig:deformation-initial}(Color online.) The initial logical
operators and stabilizer generator for the linear logical Majorana array.}
\end{figure}

To begin the deformation, introduce a second line of logical Majorana fermions $c_{2n+2}...c_{4n+3}$
prepared in the $+1$ eigenstate of a new set of stabilizer generators
\begin{align}
S_k &= c_{(2n+2)+2k} c_{(2n+2)+(2k+1)} \\
S_F &= \prod_{i=2n+2}^{4n+3} c_{i}.
\end{align}
By multiplying by these stabilizer generators, we can rewrite our logical operator generators as
\begin{align}
\begin{split}
L_k \rightarrow {}& L_k \prod_{i=0}^{k} S_i \\
= {}& \prod_{i=0}^{k} c_{2i} c_{2i+1} c_{(2n+2)+2i} c_{(2n+2)+(2i+1)}
\end{split}
\end{align}
(with no change to the $L'_k$).
\begin{figure}[H]
\center{
  \includegraphics[height=0.65in]{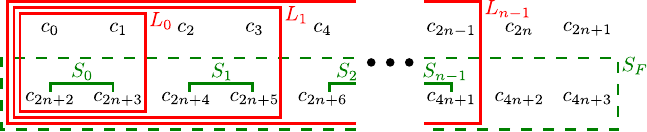}
}
\caption{\label{fig:deformation-extend}(Color online.) The second line of
logical Majorana fermions, new
stabilizer generators, and extended logical operators produced.}
\end{figure}

We can then, one at a time from $k=n-1$ to $k=0$, replace each $S_k$ with the
anticommuting stabilizer generator
\begin{align}
\tilde{S}_k = c_{2k+1} c_{2k+2} c_{(2n+2)+(2k+1)} c_{(2n+2)+(2k+2)}.
\end{align}
Note that, if the replacements are done in descending order of \(k\), the new
stabilizer generator will commute with all of our other existing stabilizer and
logical operators.

Next, take the stabilizer generator $S_F$ and multiply it by the product of all
the other stabilizer generators, rewriting it as
\begin{align} \begin{split}
S_F \rightarrow {}& {-S_F} S_G \prod_{k=0}^{n-1} \tilde{S}_k \\ = {}& c_0
c_{2n+1} c_{2n+2} c_{4n+3}.
\end{split} \end{align}
Take $S_G$ and replace it with the anticommuting stabilizer generator
\begin{align}
\tilde{S}_G = c_{2n+1} c_{4n+3}.
\end{align}
As with our other replacements, this commutes with every stabilizer and logical
operator generator except for the old $S_G$ it replaces.
\begin{figure}[H]
\center{
  \includegraphics[height=0.525in]{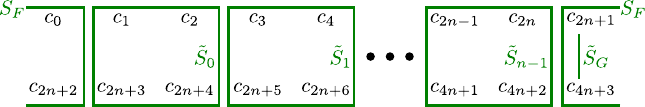}
}
\caption{\label{fig:deformation-replace}(Color online.) The stabilizer
generators after rewriting $S_F$ and replacing $S_k$ and $S_G$.}
\end{figure}

If we then multiply $S_F$ by the new
$\tilde{S}_G$, yielding
\begin{align}
S_F \rightarrow S_F \tilde{S}_G = c_0 c_{2n+2},
\end{align}
we find that $\tilde{S}_G$ is the
\emph{only} stabilizer generator supported on either $c_{2n+1}$ or $c_{4n+3}$; accordingly, we
can measure those two logical Majorana fermions off, removing $\tilde{S}_G$ entirely.
\begin{figure}[H]
\center{
  \includegraphics[height=0.5in]{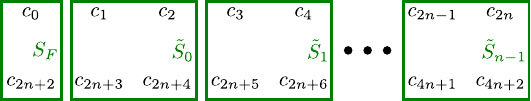}
}
\caption{\label{fig:deformation-measure}(Color online.) The stabilizer generators after
rewriting $S_F$ again and measuring off $c_{2n+1}$ and $c_{4n+3}$.}
\end{figure}

Finally, we can rewrite each $L_k$ as the
equivalent-up-to-stabilizer-group-multiplication
\begin{align}
\begin{split}
L_k \rightarrow {}& L_k S_F \prod_{i=0}^{k-1} \tilde{S}_k \\
= {}& c_{2k+1} c_{(2n+2)+(2k+1)},
\end{split}
\end{align}
leaving $S_F$ as the only stabilizer generator supported on $c_0$ and $c_{2n+2}$.  We
measure it off just as we did $\tilde{S}_G$, leaving us with the $n$ tetron
stabilizer generators $\tilde{S}_k$. Note that, as expected, each $L_k$ still anticommutes
with the corresponding $L'_k$, which has been preserved unchanged from our
original code.
\begin{figure}[H]
\center{
  \includegraphics[height=0.5in]{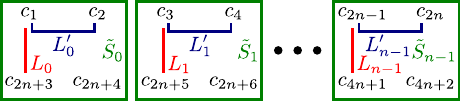}
}
\caption{\label{fig:deformation-final}(Color online.) The stabilizer generators
in the final tetron code after the deformation is complete.}
\end{figure}

Thus, we have performed the desired code deformation, from $2n+2$ logical
Majorana fermions and one weight-$(2n+2)$ stabilizer generator to $4n$ logical
Majorana fermions with $n$ weight-4 stabilizer generators, and all our logical
information preserved. Of course, this whole process can be performed in
reverse, exchanging preparations and measurements, to deform the code in the
other direction. Furthermore, any logical Majorana fermionic operator that was
even-weight (globally parity-preserving) on the original code is now even-weight
on \emph{each patch} of the new code (locally parity-preserving). Thus, any such
operator could, if desired, be measured via \emph{standard} (albeit potentially
wildly nonlocal) lattice surgery operations.

But what did it cost us? Our code deformation interacts with every logical
Majorana fermion, and indeed introduces a new weight-(\(2n+2\)) stabilizer
generator. In other words, it is as nonlocal as a code deformation could
possibly be; we should not find this surprising, because we needed to take the
extremely nonlocal
\begin{align}
L_{n-1} = \prod_{i=0}^{n-1} c_{2i} c_{2i+1}
\end{align}
to a completely local (weight-2) operator.
%
%This seems a bit much: It would be shocking and implausible if that operation
%\emph{could} be performed with entirely local operations!
%
If one was hoping, by using tetron surface-code patches instead of one giant
patch, that logical Majorana operations would be more local, or easier to route,
or some similar benefit over nonlocal logical Jordan-Wigner operations, such
benefits disappear when accounting for the the cost of moving between the
many-tetron and one-giant-patch encoding.

%%%%%%%%%%%%%%%%%%%%%%%%%%%%%%%%%%%%%%%%%%%%%%%%%%%%%%%%%%%%%%%%%%%%%
% Section
%
\section{Stride-$k$ unary iteration circuits}
\label{sec:stride-unary-iteration}

While constructing the block-encoding circuits in Sec.~\ref{sec:qubitized-hubbard}, we
discovered that we required a slightly different construction of the unary-iterated
Majorana operators \cite{Gidney:2019} than that used by Babbush
\etal~\cite{Babbush:2018a}. We refer to this construction as \emph{stride} unitary
iteration by analogy to the stride parameter present
in many classical programming languages' iteration constructs.

Conventional unary iteration circuits iterate over each individual qubit
in a target register, with the index specifying which qubit to apply a gate to, or where
to start/stop applying a Pauli string or similar gate to every qubit iterated over. We
introduce a stride parameter $k$, allowing a unary iteration circuit to iterate over only
$1/k$ of the qubits in a register (at $1/k$ the $T$-count) while still applying Pauli
strings across every qubit in the register.

%Additionally, we will include an index in our unary iteration ($0$ in this diagram, but
%labeled more generally as \(\varnothing\) elsewhere in the paper) that represents a no-op;
%this is more efficient than adding a second control qubit to enable or disable the unary
%iteration, at least if the number of indices is not already a power of 2.

Additionally, we will include an index in our unary iteration that, instead of
corresponding to a target qubit, corresponds to applying the operation to no qubits
(that is, it corresponds to a \textsc{noop}). This is more efficient than adding a second control
qubit to enable or disable the unary iteration, if the number of indices is not
already a power of 2. We will label this index $\varnothing$.

Below, we construct example circuits for $_{(2)}\mathrm{In}_l$, the stride-2 unary
iteration operator, applying Jordan-Wigner encoded Majorana operators on a register
$\psi$ of six qubits. These qubits are grouped as pairs addressed by
$l \in \left\{{01}, {10}, {11}\right\}$, one representing the spin-up fermion and
the other representing the spin-down fermion on each lattice site. Additionally,
when $\ket{l} = \ket{00}$ the circuit performs \textsc{noop}; that is, we define
$\varnothing = 00$.

For the spin-up case, \(_{(2)}\mathrm{In}_l \left(\vec{Z} Y_{l,\uparrow}\right)\), we have

\onecolumngrid

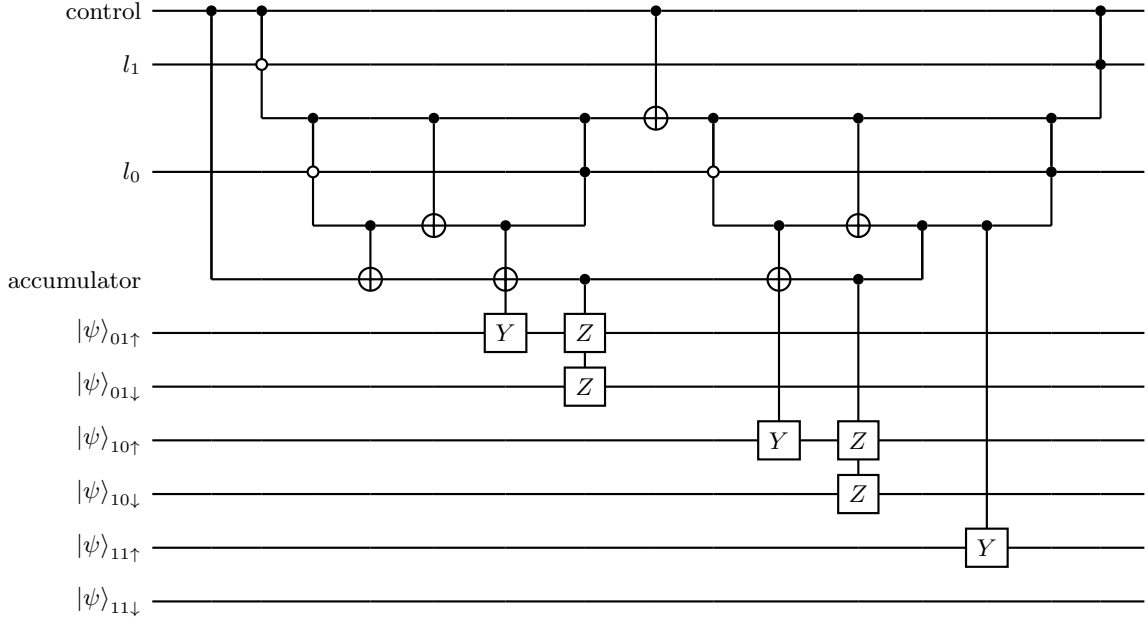
\begin{figure}[H]
\begin{center}
\resizebox{0.85\textwidth}{!}{%
\begin{quantikz}[row sep={0.7cm,between origins}]
\lstick{${\mathrm{control}}$} &[2mm] \ctrl{5} & \ctrl{1} & \qw & \qw & \qw & \qw & \qw & \ctrl{2} & \qw & \qw & \qw & \qw & \qw & \qw & \ctrl{1} & \qw \\
\lstick{${l_1}$} &[2mm] \qw & \octrl{-1} & \qw & \qw & \qw & \qw & \qw & \qw & \qw & \qw & \qw & \qw & \qw & \qw & \ctrl{-1} & \qw \\
&[2mm] & \vqw{-1} & \ctrl{1} & \qw & \ctrl{2} & \qw & \ctrl{1} & \targ{} & \ctrl{1} & \qw & \ctrl{2} & \qw & \qw & \ctrl{1} & \qw\vqw{-1} \\
\lstick{${l_0}$} &[2mm] \qw & \qw & \octrl{-1} & \qw & \qw &\qw & \ctrl{-1} & \qw & \octrl{-1} & \qw & \qw & \qw & \qw & \ctrl{-1} & \qw & \qw \\
&[2mm] & & \vqw{-1} & \ctrl{1} & \targ{} & \ctrl{2} & \qw\vqw{-1} & & \vqw{-1} & \ctrl{4} & \targ{} & \ctrl{1} & \ctrl{6} & \qw\vqw{-1} \\
\lstick{${\mathrm{accumulator}}$} &[2mm] \vqw{-5} & \qw & \qw & \targ{} & \qw & \targ{} & \ctrl{2} & \qw & \qw & \targ{} & \ctrl{4} & \qw\vqw{-1} & \\
\lstick{${\ket{\psi}_{01\uparrow}}$} &[2mm] \qw & \qw & \qw & \qw & \qw & \gate{Y} & \gate{Z} & \qw & \qw & \qw & \qw & \qw & \qw & \qw & \qw & \qw \\
\lstick{${\ket{\psi}_{01\downarrow}}$} &[2mm] \qw & \qw & \qw & \qw & \qw & \qw & \gate{Z} & \qw & \qw & \qw & \qw & \qw & \qw & \qw & \qw & \qw \\
\lstick{${\ket{\psi}_{10\uparrow}}$} &[2mm] \qw & \qw & \qw & \qw & \qw & \qw & \qw & \qw & \qw & \gate{Y} & \gate{Z} & \qw & \qw & \qw & \qw & \qw \\
\lstick{${\ket{\psi}_{10\downarrow}}$} &[2mm] \qw & \qw & \qw & \qw & \qw & \qw & \qw & \qw & \qw & \qw & \gate{Z} & \qw & \qw & \qw & \qw & \qw \\
\lstick{${\ket{\psi}_{11\uparrow}}$} &[2mm] \qw & \qw & \qw & \qw & \qw & \qw & \qw & \qw & \qw & \qw & \qw & \qw & \gate{Y} & \qw & \qw & \qw \\
\lstick{${\ket{\psi}_{11\downarrow}}$} &[2mm] \qw & \qw & \qw & \qw & \qw & \qw & \qw & \qw & \qw & \qw & \qw & \qw & \qw & \qw & \qw & \qw \\
\end{quantikz}
}%
\caption{\label{fig:unary-majorana-up}A spin-up Majorana fermion operator, Jordan-Wigner
encoded, and implemented using stride-2 unary iteration.}
\end{center}
\end{figure}

Similarly, for the spin-down
\(_{(2)}\mathrm{In}_l \left(\vec{Z} Y_{l,\downarrow}\right)\), we have the circuit
\begin{figure}[H]
\begin{center}
\resizebox{0.85\textwidth}{!}{%
\begin{quantikz}[row sep={0.7cm,between origins}]
\lstick{${\mathrm{control}}$} &[2mm] \ctrl{5} & \ctrl{1} & \qw & \qw & \qw & \qw & \qw & \ctrl{2} & \qw & \qw & \qw & \qw & \qw & \qw & \ctrl{1} & \qw \\
\lstick{${l_1}$} &[2mm] \qw & \octrl{-1} & \qw & \qw & \qw & \qw & \qw & \qw & \qw & \qw & \qw & \qw & \qw & \qw & \ctrl{-1} & \qw \\
&[2mm] & \vqw{-1} & \ctrl{1} & \qw & \ctrl{2} & \qw & \ctrl{1} & \targ{} & \ctrl{1} & \qw & \ctrl{2} & \qw & \qw & \ctrl{1} & \qw\vqw{-1} \\
\lstick{${l_0}$} &[2mm] \qw & \qw & \octrl{-1} & \qw & \qw &\qw & \ctrl{-1} & \qw & \octrl{-1} & \qw & \qw & \qw & \qw & \ctrl{-1} & \qw & \qw \\
&[2mm] & & \vqw{-1} & \ctrl{1} & \targ{} & \ctrl{3} & \qw\vqw{-1} & & \vqw{-1} & \ctrl{5} & \targ{} & \ctrl{1} & \ctrl{7} & \qw\vqw{-1} \\
\lstick{${\mathrm{accumulator}}$} &[2mm] \vqw{-5} & \qw & \qw & \targ{} & \ctrl{1} & \targ{} & \ctrl{3} & \qw & \qw & \targ{} & \ctrl{5} & \qw\vqw{-1} & \\
\lstick{${\ket{\psi}_{01\uparrow}}$} &[2mm] \qw & \qw & \qw & \qw & \gate{Z} & \qw & \qw & \qw & \qw & \qw & \qw & \qw & \qw & \qw & \qw & \qw \\
\lstick{${\ket{\psi}_{01\downarrow}}$} &[2mm] \qw & \qw & \qw & \qw & \qw & \gate{Y} & \gate{Z} & \qw & \qw & \qw & \qw & \qw & \qw & \qw & \qw & \qw \\
\lstick{${\ket{\psi}_{10\uparrow}}$} &[2mm] \qw & \qw & \qw & \qw & \qw & \qw & \gate{Z} & \qw & \qw & \qw & \qw & \qw & \qw & \qw & \qw & \qw \\
\lstick{${\ket{\psi}_{10\downarrow}}$} &[2mm] \qw & \qw & \qw & \qw & \qw & \qw & \qw & \qw & \qw & \gate{Y} & \gate{Z} & \qw & \qw & \qw & \qw & \qw \\
\lstick{${\ket{\psi}_{11\uparrow}}$} &[2mm] \qw & \qw & \qw & \qw & \qw & \qw & \qw & \qw & \qw & \qw & \gate{Z} & \qw & \qw & \qw & \qw & \qw \\
\lstick{${\ket{\psi}_{11\downarrow}}$} &[2mm] \qw & \qw & \qw & \qw & \qw & \qw & \qw & \qw & \qw & \qw & \qw & \qw & \gate{Y} & \qw & \qw & \qw \\
\end{quantikz}
}%
\end{center}
\caption{\label{fig:unary-majorana-down}A spin-down Majorana fermion operator, Jordan-Wigner
encoded, and implemented using stride-2 unary iteration.}
\end{figure}
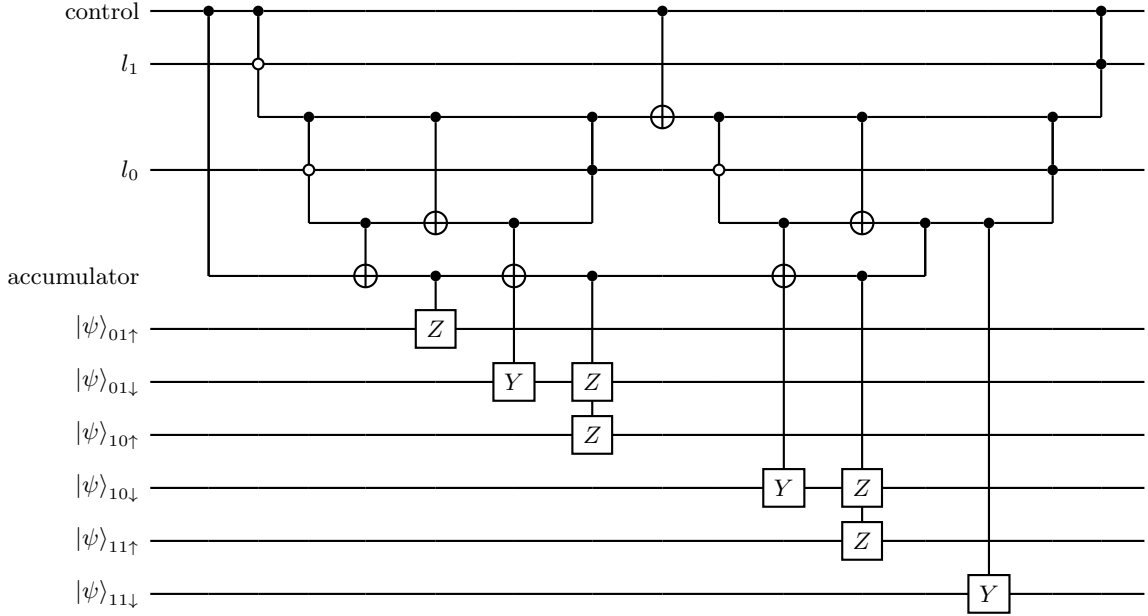
\vspace{0\baselineskip}
\twocolumngrid

This has a \(T\)-cost of \(4 L - 4\), just as for standard unary iteration, but here
\(L = \frac{N}{2} + 1\) for a \(T\)-count of \(2 N\) in contrast to the full
(stride-1) indexed Majorana operator's \(4 N - 4\).